\documentclass[pra,aps,showpacs,floatfix,twocolumn,nofootinbib]{revtex4-1}
\usepackage{amsmath}
\usepackage{amsfonts}
\usepackage{amssymb}
\usepackage{revsymb}
\usepackage{graphicx}
\usepackage{siunitx}
\usepackage{bm}
\usepackage{dcolumn}
\usepackage{multirow}
\usepackage[version=3]{mhchem}

\newcommand{\eref}[1]{Eq.~(\ref{#1})}
\newcommand{\tref}[1]{Table~\ref{#1}}

\def\eval#1{\left\langle #1 \right\rangle}

\begin{document}
\title{Atomic Properties of Lu$^+$.}
\author{Eduardo Paez}
\affiliation{Center for Quantum Technologies, 3 Science Drive 2, Singapore, 117543}
\affiliation{Department of Physics, National University of Singapore, 2 Science Drive 3, Singapore, 117551}
\author{K. J. Arnold}
\affiliation{Center for Quantum Technologies, 3 Science Drive 2, Singapore, 117543}
\affiliation{Department of Physics, National University of Singapore, 2 Science Drive 3, Singapore, 117551}
\author{Elnur Hajiyev}
\affiliation{Center for Quantum Technologies, 3 Science Drive 2, Singapore, 117543}
\affiliation{Department of Physics, National University of Singapore, 2 Science Drive 3, Singapore, 117551}
\author{M. D. Barrett}
\email{phybmd@nus.edu.sg}
\affiliation{Center for Quantum Technologies, 3 Science Drive 2, Singapore, 117543}
\affiliation{Department of Physics, National University of Singapore, 2 Science Drive 3, Singapore, 117551}
 \author{S.~G.~Porsev$^{1,2}$}
 \author{V.~A.~Dzuba$^3$}
 \author{U.~I.~Safronova$^{4}$}
 \author{M.~S.~Safronova$^{1,5}$}
 \affiliation{
$^1$Department of Physics and Astronomy, University of Delaware, Newark, Delaware 19716, USA\\
$^2$Petersburg Nuclear Physics Institute, Gatchina, Leningrad District, 188300, Russia,\\
$^3$School of Physics, University of New South Wales, Sydney 2052, Australia, \\
$^4$Physics Department, University of Nevada, Reno, Nevada 89557, USA\\
$^5$Joint Quantum Institute, National Institute of Standards and Technology and the University of Maryland,
College Park, Maryland, 20742, USA}

\begin{abstract}
Singly ionised Lutetium has recently been suggested as a potential clock candidate.  Here we report a joint experimental and theoretical investigation of \ce{Lu^+}.    Measurements relevant to practical clock operation are made and compared to atomic structure calculations.  Calculations of scalar and tensor polarizabilities for clock states over a range of wavelengths are also given.  These results will be useful for future work with this clock candidate.
\end{abstract}

\maketitle
\section{Introduction}
The development of atomic clocks has played an important role in todays society with applications in many different technologies, most notably the Global Positioning System and navigation.  Increased levels of performance have allowed tests of fundamental physics \cite{alpha} and new avenues of exploration in quantum many body physics \cite{SUN, QMBYe}. Increasing levels of accuracy and stability continue to be made with atomic clocks based on optical transitions in isolated atoms \cite{AlIon,SrYe,HgIon,SrIon,YbIon,InIon,YbForbidden,RMP,Ludlow}.  By now a number of groups have demonstrated superior performance over the current caesium frequency standards with the best clocks to date having inaccuracy at the $10^{-18}$ level \cite{AlIon,SrYe,SrYe2}.  For ion-based clocks, a significant bottleneck to improved levels of accuracy is the relatively low stability achieved with a single ion.  Recently singly ionised Lutetium has been proposed as a possible candidate to overcome this hurdle \cite{MDB1,MDB2}.  

The clock transition in singly ionised Lutetium is a highly forbidden $M1$ \ce{^1$S$_0} to \ce{^3$D$_1} transition \cite{MDB1,Dzuba}.  This ion has a number of fortuitous properties that are almost ideally suited for clock applications \cite{MDB1,MDB2}.  The $2.45\,\mathrm{MHz}$ linewidth of the \ce{^3$D$_1} to \ce{^3$P$^o_0}  detection transition provides the possibility of a very low Doppler cooling limit and yet sufficiently large for practical detection.  A novel averaging scheme eliminates shifts associated with the $J=1$ level placing it on an equal footing with $J=0$ to $J=0$ candidates \cite{MDB1}.    A very large hyperfine and fine structure splitting results in a very low magnetic field dependence of both the average frequency and the component transitions contributing to the average.  Finally, initial estimates of the differential scalar polarisability indicate that it is sufficiently small to allow practical room temperature operation, with a sign that allows micromotion shifts to be eliminated. This latter property has kindled the idea of clock operation on large ion crystals \cite{MDB2}.

All of the available low-lying $D$ states in \ce{Lu^+} are potentially long lived.  These spectator states could be in principle be used as clock states themselves.  However, in so far as clock operation with the \ce{^3$D$_1} state is concerned, the remaining $D$ levels could potentially complicate clock operation via the need for a more complicated laser system. In this paper we give a detailed investigation of these potential issues using \ce{^{175}Lu^+}.  Measurements of lifetimes and branching ratios relevant to practical clock operation are made and compared to atomic structure calculations.  In addition we provide calculations of scalar and tensor polarizabilities for clock states over a range of wavelengths.  This work provides the first step in evaluating the potential of this clock candidate and the calculations given will provide a useful reference for future experimental work.

\section{Experiment Setup}
\label{Setup}
\subsection{Apparatus}
The experiments are performed in a four-rod linear Paul trap with axial end caps, similar to the ones described in \cite{BoonLeng2,Nick}. The trap consists of four stainless steel rods of diameter $1.0\,\mathrm{mm}$ whose centers are arranged on the vertices of a square with $3.6\,\mathrm{mm}$ length of the side. A $3.6\,\mathrm{MHz}$ rf potential is applied via a step-up transformer to two diagonally opposing electrodes. A small DC voltage applied to the other two electrodes ensures a splitting of the transverse trapping frequencies. Axial confinement is provided by two axial pins separated by $7\,\mathrm{mm}$. Using this configuration, the measured trapping frequencies are $(\omega_x, \omega_y, \omega_z)/2\pi \approx (350, 300, 80)\,\mathrm{kHz}$.  These frequencies were measured using \ce{^{138}Ba^+} which is used throughout for continuous sympathetic cooling.

\begin{figure}
\begin{center}
\includegraphics[width=8cm]{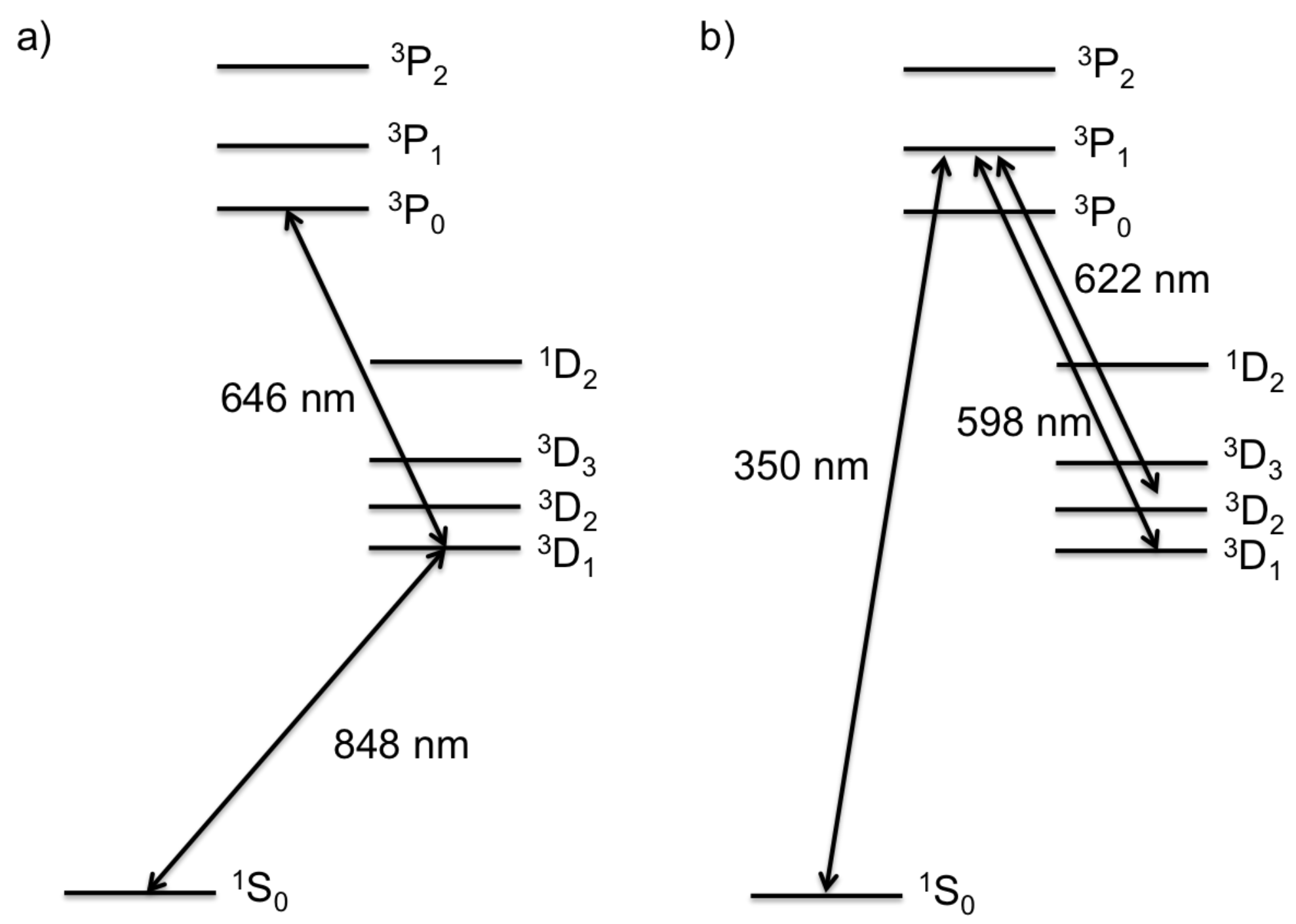}
\caption{\ce{Lu^+} level structure showing  (a) the 848 nm clock and 646 nm detection transitions and (b) the repumping lasers used to optically pump into and out of the \ce{^3$D$_1} level.}
\label{LuStructure}
\end{center}
\end{figure}

The level structure for Lu$^+$ is given in Fig.~\ref{LuStructure} showing the \ce{^1$S$_0} to \ce{^3$D$_1} clock transition, and the \ce{^3$D$_1} to \ce{^3$P$^o_0} transition for detection and cooling.  Optical pumping and state preparation is achieved via the \ce{^3$P$^o_1} level.  The experiments reported here use \ce{^{175}Lu^+} which has a nuclear spin $I=7/2$.  The 350 nm laser is a frequency doubled diode and addresses the \ce{^1$S$_0} $F=7/2$ to \ce{^3$P$^o_1} $F'=7/2$ transition.  It propagates orthogonal to a \SI{0.5}{\milli\tesla} B-field and is linearly polarized along the direction of the field.  The measured optical pumping time out of the \ce{^1$S$_0} level is \SI{2}{\micro\second} which is the $1/e$ decay time of the \ce{^1$S$_0} population.  The 598 nm laser is also a frequency doubled diode laser and addresses the \ce{^3$D$_1} $F=9/2$ to \ce{^3$P$^o_1} $F'=9/2$ transition.  Optical pumping out of the \ce{^3$D$_1} level is achieved in conjunction with the 646 nm laser and the measured optical pumping time is \SI{6}{\micro\second}.  The 622 nm laser is a multimode laser which is sufficiently broad to address all hyperfine states of the \ce{^3$D$_2} to \ce{^3$P$^o_1} transition and the measured optical pumping time is \SI{10}{\micro\second}.  Both the 598 nm and 622 nm laser are linearly polarized and propagate along the B-field.
\subsection{Detection}
\label{Detection}
As shown in Fig.~\ref{LuStructure}, detection is achieved via scattering on the \ce{^3$D$_1} to  \ce{^3$P$^o_0} levels.  To address the three separate hyperfine levels, a wideband electro-optic modulator (EOM) generates sidebands of approximately \SI{8}{\giga\hertz} which are separated from the carrier using a cavity. The carrier is frequency shifted via an acousto-optic modulator (AOM) before being recombined with the sidebands.  This provides independent frequency control of all three beams.  All beams are linearly polarized and propagate along the \SI{0.5}{\milli\tesla} B-field.

Fluorescence at \SI{646}{\nano\meter} is collected onto an avalanche photodiode (APD).  A narrowband filter eliminates scattered light from all other light sources including the \SI{650}{\nano\meter} light used for cooling \ce{^{138}Ba^+}.  This allows continuous sympathetic cooling throughout the \ce{^{175}Lu^+} detection window.  Since the ion is continuously cooled, we can operate at near full saturation for optimum detection efficiency and we typically achieve a mean photon count rate of $\gtrsim 5$ photons/ms.

For the experiments reported here, we desire a detection scheme to determine when the ion goes bright (dark) with high detection efficiency.  To do this we use a Bayesian detection scheme similar to that reported in \cite{Lucas}.  From the number of photons collected in a given detection time step, we update the probability that the ion is in a bright state via
\begin{equation}
\label{Bayesian}
P(\mathrm{b}|n)=\frac{P(n|\mathrm{b})P(b)}{P(n|\mathrm{b})P(b)+P(n|\mathrm{d})P(d)},
\end{equation}
where $P(b|n)$ ($P(d|n)$) is the conditional probability the ion is in a bright (dark) state given $n$ photons, $P(n|b)$ ($P(n|d)$) is the conditional probability of getting $n$ photons given the ion is in a bright (dark) state, and $P(b)$ (($P(d)$)) is the current probability the ion is in the bright (dark) state.  The probability $P(b)$ is updated in real time via a field programmable gate array with the conditional probabilities $P(n|b)$ and $P(n|d)$ stored on chip.  Detection is initiated with $P(b)=0.5$ and terminated when $P(b)$ reaches pre-programmed thresholds for bright and dark states.  We note that the performance of this scheme is insensitive to the choice of time step.

When continuously monitoring for a state change, $P(b)$ is initialised to 0.5 and updated in subsequent detection windows to $P(b|n)$ according to eq~\ref{Bayesian}.  If $P(b)$ falls below (above) 0.5, the ion is assumed to have gone dark (bright) and detection continues until the appropriate threshold is reached, in which case the state change is deemed verified.  Alternatively, if $P(b)$ subsequently falls above (below) 0.5, the state change is deemed in error, and $P(b)$ is reinitialised to $0.5$.  The error rate for determining the ion is in the bright state is limited by hyperfine induced decay from \ce{^3$P$^o_0} to states other than \ce{^3$D$_1}.  Thus $P(b)$ cannot achieve values arbitrarily close to one before the ion decays to a dark state.  Similarly, determination of the \ce{^3$D$_2} dark state is limited by possible decay to the \ce{^3$D$_1}. 

For the experiments reported in section~\ref{3P0Measurements}, we also require an accurate estimate of the total collection efficiency.  This is achieved using \ce{^{138}Ba^+} by repeated cycles of optical pumping between the \ce{S_{1/2}} and \ce{D_{3/2}} levels.  Optically pumping from \ce{S_{1/2}} to \ce{D_{3/2}}, produces precisely one \SI{650}{\nano\meter} photon.  From the photons collected over several million cycles we infer a collection efficiency of 0.00326(2).
\section{Measurements}
\subsection{\ce{^3$P$^o_1} branching ratios}
\label{3P1Measurements}
Optical pumping via the \ce{^3$P$^o_1} level results in undesired population of the \ce{^1$D$_2} metastable level.  The \ce{^3$P$^o_1} level decays to \ce{^1$S$_0}, \ce{^3$D$_1}, \ce{^3$D$_2} and \ce{^1$D$_2} with respective decay rates $W_0, W_1, W_2,$ and $W_3$ and branching ratios $B_k=W_k/\sum W_k$.  Since $\sum B_k=1$ we need three more equations to uniquely determine $W_k$.  This is achieved via three separate optical pumping experiments.

We first prepare the ion in \ce{^3$D$_1} by optically pumping with the $350, 622,$ and $646$ nm lasers until the ion is bright.  For this step, we set the threshold count rate to a high value to ensure the initial state is bright with high probability.  We then optically pump the ion into \ce{^1$S$_0} (\ce{^3$D$_2}) using the 646, 598 and 622 (350) nm lasers.  The ion is then pumped out of the \ce{^1$S$_0} (\ce{^3$D$_2}) level using the 350 (622) nm laser and the population, $P_0$ ($P_1$), in \ce{^3$D$_1} is measured.  Neglecting any decay of population appearing in \ce{^1$D$_2} we have
\begin{equation}
P_0=\frac{B_0}{B_0+B_3}\frac{B_1}{1-B_0},\quad P_1=\frac{B_2}{B_2+B_3}\frac{B_1}{1-B_2}
\end{equation}
Similarly, optical pumping to \ce{^3$D$_2}, followed by optical pumping with both the 350, and 622 nm lasers, gives a population, $P_2$, in the \ce{^3$D$_1} level of
\begin{equation}
P_2=\frac{B_2}{B_2+B_3}\frac{B_1}{1-B_0-B_2}.
\end{equation}
For each $P_k$, $2\times 10^4$ measurements were made giving $P_0=0.3027(32)$, $P_1=0.3166(33)$ and $P_2=0.9669(13)$. The inferred branching ratios from these measurements are given in table~\ref{3P1} along with theoretical estimates from section~\ref{Sect_lifetimes}.
\begin{table}
\caption{Branching ratios for decay from \ce{^3$P$^o_1}.  Theoretical values are from calculations given in section~\ref{Sect_lifetimes}.}
\begin{ruledtabular}
\centering
\begin{tabular}{l c c}
Lower level & Exp. & Theory \\
\hline
$6s^2$ \ce{^1$S$_0} & 0.3915(44) & 0.376\\
$6s5d$ \ce{^3$D$_1} & 0.1862(17) & 0.186 \\
$6s5d$ \ce{^3$D$_2} & 0.4178(45) & 0.435 \\
$6s5d$ \ce{^1$D$_2} & 0.00438(18) & 0.0036 
\end{tabular}
\label{3P1}
\end{ruledtabular}
\end{table}
The error bars given are the statistical error.  The main systematic is due to decay of the \ce{^1$D$_2} during optical pumping.  Since the measured optical pumping times for each laser is $\sim 10\,\mathrm{\mu s}$, which is much less than the  \ce{^1$D$_2} lifetime as discussed in the next section, the effect of the decay is less than the statistical error.

There is fair agreement between the experimental and theoretical results with the three main decay channels being within $4\%$.  The larger discrepancy of $\sim18\%$ for decay to \ce{^1$D$_2} can be expected given the significantly smaller decay rate. 
\subsection{\ce{^1$D$_2} Lifetime}
\label{Sect1D2}
\begin{figure}
\begin{center}
\includegraphics[width=8cm]{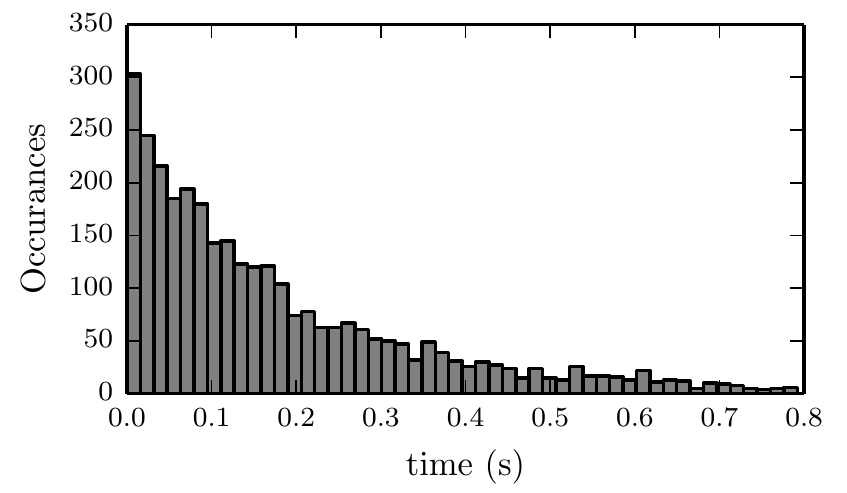}
\caption{Histogram of dark times associated with the \ce{^1$D$_2} decay.  We have omitted all times less than \SI{200}{\milli\second} or greater than \SI{1}{\second} with \SI{200}{\milli\second} subtracted of the remaining times.}
\label{1D2fig}
\end{center}
\end{figure}

To measure the  \ce{^1$D$_2} lifetime, we first optically pump to this level using the 350, 598, 622, and \SI{646}{nm} lasers.   After (\SI{10}{ms}), we switch off the 598 nm laser and monitor fluorescence of the 646 nm light.  The \ce{^1$D$_2} lifetime is due to an $E2$ decay to \ce{^1$S$_0}.  However, spin mixing gives a small contribution from $M1$ transitions to \ce{^3$D$_$2$} as discussed in section~\ref{Sect_lifetimes} and values of relevant transitions are tabulated in table~\ref{branching}.  Decay to \ce{^3$D$_3} occurs with a branching ratio $\sim 1\%$ and the lifetime of this state is $>10\,\mathrm{s}$.  Hence, these decays are infrequent and result in very long dark periods.  Decay to \ce{^3$D$_2} or \ce{^1$S$_0}  occurs with probability $q_S$.  These levels are optically pumped to the detection level, \ce{^3$D$_1}, with a small probability, $p_S$, of being repumped back to \ce{^1$D$_2}.  Neglecting effects of decays to \ce{^3$D$_3} and optical pumping times, the distribution of dark times is exponential with a rate $W_S^{(m)}=(1-p_Sq_S)W_S$ where $W_S$ is the total linewidth of the \ce{^1$D$_2} level.  In Fig.\ref{1D2fig}, we give the measured distribution of dark times from which we infer $W_S^{(m)}=5.41(12)\,\mathrm{s}^{-1}$.  For this data we have eliminated all times less than \SI{200}{\milli\second} or greater than \SI{1}{\second} with \SI{200}{\milli\second} subtracted of the remaining times.  Eliminating times less than \SI{200}{\milli\second} removes any data points resulting from imperfect optical pumping to the \ce{^1$D$_2} level, and eliminating all times greater than \SI{1}{\second} eliminates a small number of events associated with decay into \ce{^3$D$_3}.
Denoting the $M1$ decay rates from \ce{^1$D$_2} to \ce{^3$D$_$2$} by $W_{S,J}$ and the $E2$ decay rate from \ce{^1$D$_2} to \ce{^1$S$_0} by $W_{S,0}$, we can express the total linewidth, $W_S$, by
\begin{equation}
\label{E2AS}
W_{S,0}=\frac{W_S^{(m)}-(W_{S,1}+W_{S,3})}{1-p_S}-W_{S,2},
\end{equation}
and
\begin{equation}
W_S= \frac{W_S^{(m)}}{1-p_S}-\frac{p_S}{1-p_S}(W_{S,1}+W_{S,3}),
\end{equation} 
where we have used the fact that
\begin{equation}
q_S=\frac{W_{S,0}+W_{S,2}}{W_S}.
\end{equation}
In terms of the branching ratios, $B_k$, from section~\ref{3P1Measurements}, we have
\begin{equation}
p_S=\frac{B_3}{B_1+B_3}.
\end{equation}
This maybe determined from the measurements, $P_k$, made in section~\ref{3P1Measurements} and we infer $p_S=0.02297(88)$.  From the calculated $M1$ transition rates given in table~\ref{branching} we infer decay rates $5.20(12)\,\mathrm{s}^{-1}$ and $5.53(12)\,\mathrm{s}^{-1}$ for $W_{S,0}$ and $W_S$ respectively.  The errors given include only the statistical uncertainty from the experimental measurements.  We note that the measured rates are $\sim30\%$ larger than the theoretical estimates given in section~\ref{Sect_lifetimes}. 

\subsection{\ce{^3$D$_2} Lifetime}
\begin{figure}
\begin{center}
\includegraphics[width=8cm]{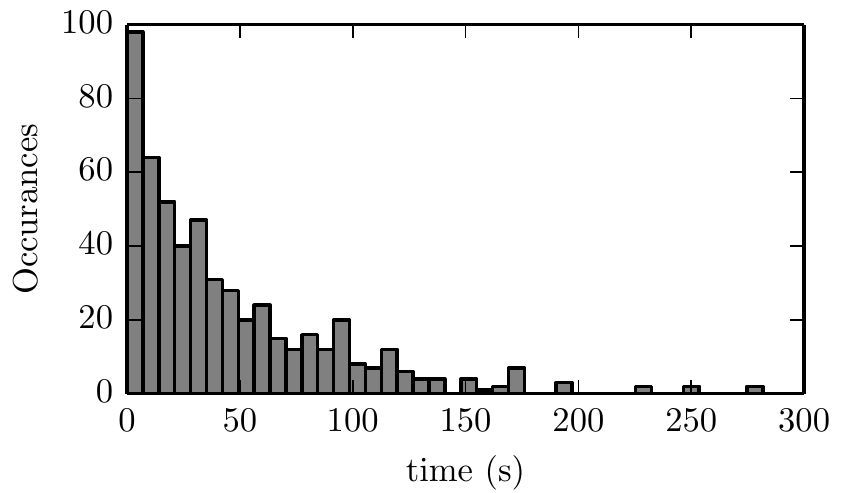}
\caption{Histogram of dark times associated with the \ce{^3$D$_2} decay.}
\label{3D2fig}
\end{center}
\end{figure}

We measure the \ce{^3$D$_2} lifetime similar to the \ce{^1$D$_2} case.  We first optically pump to \ce{^3$D$_2} using the 350, 598, and 646 nm lasers.   After $(10\,\mathrm{ms})$, we switch off the 598 nm laser and monitor fluorescence of the 646 nm light.  The \ce{^3$D$_2} lifetime is due to a spin forbidden $E2$ decay to \ce{^1$S$_0} with a small contribution from an $M1$ decay to \ce{^3$D$_1}.  Decays to \ce{^1$S$_0} result in optical pumping to \ce{^3$D$_1} and repumping to \ce{^3$D$_2}.  Neglecting optical pumping times, the distribution of dark times is also exponential with  a rate $W_{T}^{(m)}=(1-p_Tq_T)W_T$, where $p_T$ is the probability of being repumped from \ce{^1$S$_0} to \ce{^3$D$_2} and $q_T$ is the branching ratio for decay from \ce{^3$D$_2} to \ce{^1$S$_0}.  In Fig.\ref{3D2fig}, we give the measured distribution of dark times from which we infer $W_T^{(m)}=0.022(1)\,\mathrm{s}^{-1}$. Note that, for each dark cycle, optically pumping to the \ce{^3$P$^o_1} can result in population of \ce{^1$D$_2} which extends the optical pumping time.  Since the probability that this occurs is small and the lifetime of this state is much less than the measured mean dark time we may neglect this effect. 

We can express the $E2$ decay rate, $W_{T,0}$, and total linewidth $W_T$ as
\begin{equation}
\label{E2AT}
W_{T,0}=\frac{1}{1-p_T}\Big(W_T^{(m)}-W_{2,1}\Big)
\end{equation}
and
\begin{equation}
W_T= \frac{W_T^{(m)}}{1-p_T}-\frac{p_T}{1-p_T}W_{2,1}
\end{equation} 
where $W_{2,1}$ is the $M1$ decay rate for the \ce{^3$D$_2} to \ce{^3$D$_1} transition and we have used the fact that
\begin{equation}
q_T=\frac{W_{T,0}}{W_T}.
\end{equation}
In terms of the branching ratios, $B_k$, from section~\ref{3P1Measurements}, we have
\begin{equation}
p_T=\frac{B_2}{1-B_0}.
\end{equation}
This maybe determined from the measurements, $P_k$, made in section~\ref{3P1Measurements} and we infer $p_T=0.6917(33)$.
Together with the calculated $M1$ transition rates given in table~\ref{branching} we infer decay rates $0.0519(33)$ and $0.0579(33)$ for $W_{T,0}$ and $W_T$ respectively.  The errors given include only the statistical uncertainty from the experimental measurements.  We note that the measured rates are $\sim25\%$ larger than the theoretical estimates given in section~\ref{Sect_lifetimes}.

Given that the measured lifetime is very long, measurements could potentially be compromised by off-resonant scattering out of \ce{^3$D$_2} by the Barium cooling lasers, the $350$ repump laser or the \SI{646}{nm} detection beam.  Of these, the most significant scattering rate is from coupling to the 5d5p \ce{^3F_2} level by the \SI{350}{nm} laser.  From dipole matrix elements given in table~\ref{Pol_stat}  and a measured intensity of $\sim$\SI{500}{\mW \per \cm^{2}}, the scattering rate from \ce{^3$D$_2} to \ce{^3$D$_1} averaged over all possible \ce{^3$D$_2} states is $\sim 3.5\times10^{-5}\,\mathrm{s}^{-1}$.  This less than $1\%$ of the calculated $M1$ decay rate between these states and so contributes much less than the statistical error to the overall decay rates.  We can expect scattering rates to \ce{^3$D$_3} and \ce{^1$D$_2} to be of a similar magnitude and thus equally negligible.

\subsection{Hyperfine quenching of \ce{^3$P$^o_0}.}
\label{3P0Measurements}
Decay from \ce{^3$P$^o_0} to \ce{^3$D$_1} is the only dipole allowed transition from \ce{^3$P$^o_0}.  However, the hyperfine interaction induces a low multipole electromagnetic decay to other states.  In the case of \ce{Lu^+}, this results in a quenching of the fluorescence rate for the \ce{^3$D$_1} to \ce{^3$P$^o_0} detection channel.  When fluorescing on this transition, the rate of scattering out of the detection channel is given by
\begin{equation}
\lambda=w \rho_\mathrm{ee}=\frac{w}{W}\frac{\eval{n}}{q\tau_D}
\end{equation}
where $W$ is the linewidth of the upper state, $w$ is the total decay rate from \ce{^3$P$^o_0} to states other than \ce{^3$D$_1}, $\eval{n}$ is the background subtracted mean number of photons collected in a time $\tau_\mathrm{D}$, and $q$ is the overall detection efficiency of the imaging system.  Measuring $\lambda$ involves determining when the ion goes dark and so the measured rate must also include the error rate in that determination. Even an error rate of $10^{-3}$ in a \SI{1}{ms} detection time would result in a significant contribution to the measured rate.   Since there is negligible probability of repumping from the dark state back to the bright state,  we can repeatedly test a dark state event to confirm the measurement similar to the approach reported in \cite{Lucas}.

\begin{figure}
\begin{center}
\includegraphics[width=8cm]{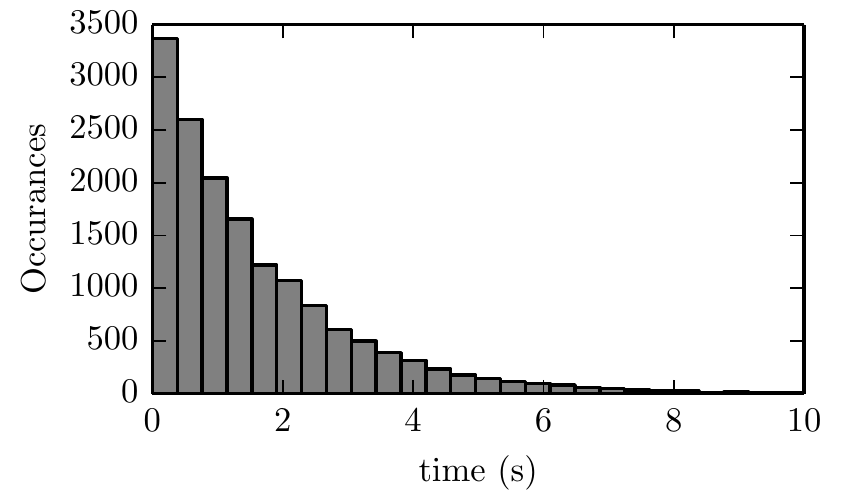}
\caption{Histogram of bright times when fluorescing on the \ce{^3$D$_1} to \ce{^3$P$^o_0} transition.}
\label{3P0fig}
\end{center}
\end{figure}

To measure $\lambda$, we first optically pump using $350, 622,$ and $646$ nm lasers until the ion is bright.  For this step, we set the threshold count rate to a high value to ensure the initial state is bright with high probability.  We then switch off the repump lasers and monitor the time the ion remains fluorescent.   The distribution of bright times is given in Fig.\ref{3P0fig} which gives a fitted value of $\lambda=0.624(5)$.  Using measured count rates for the bright and dark states of 6.290(5) and 0.560(3) per ms respectively, together with the measured detection efficiency of $0.00326(2)$ we infer a ratio $w/W=3.55(6)\times 10^{-7}$.  We used an artificial background to match the photon count rate of a bright ion to determine the contribution of the measured rate from detection errors.  Out of $10^5$ events we obtained an average detection time of 6 ms with no errors found.  This bounds the contribution to $<0.0016$/s which is well below the statistical uncertainty.

Decay out of the detection channel is dominated by decays from \ce{^3$P$^o_0} to \ce{^1$S$_0} and \ce{^3$D$_2}.  In two separate experiments, we determine the contribution from each of these decays by repumping using either 350 or 622 after the ion is confirmed dark and measure the fraction returning to the bright state.  From these measurements and the branching ratios determined in section~\ref{3P1Measurements} the percentage of decays going to \ce{^1$S$_0} and \ce{^3$D$_2} are 0.497(19) and 0.562(30) respectively.  These values are in reasonable agreement with theoretical values given in table~\ref{Qr} .

\section{Theory}
In this section we give details of atomic structure calculations.  We start with polarizabilities of relevant clock states, namely the $6s^2\,\ce{^1$S$_0}$, $5d6s\,\ce{^3$D$_1}$, and $5d6s\,\ce{^3$D$_2}$ levels.  We then determine lifetimes and branching ratios for low-lying levels, followed by a determination of the quenching rate of the \ce{^3$P$^o_0} level.
\subsection{Polarizabilities}
\label{Polar}
We evaluated the scalar static and dynamic polarizabilities of the $6s^2\,\ce{^1$S$_0}$, $5d6s\,\ce{^3$D$_1}$, and $5d6s\,\ce{^3$D$_2}$ states of
Lu$^+$ using the high-precision relativistic method that combines configuration interaction (CI) and
linearized coupled-cluster (all-order) method \cite{SafKozJoh09}.
In this CI+all-order method, the energies and wave functions are
determined from the time-independent multiparticle Schr\"odinger equation
\begin{equation}
H_{\rm eff}(E_n) \Phi_n = E_n \Phi_n,
\label{Heff}
\end{equation}
where the effective Hamiltonian is defined as
\begin{equation}
H_{\rm eff}(E) = H_{\rm FC} + \Sigma(E).
\end{equation}
Here $H_{\rm FC}$ is the Hamiltonian in the frozen core approximation and $\Sigma$ is the energy-dependent correction, which takes into account virtual core excitations in all orders. To establish the importance of the higher-order corrections, we also carried out the calculations constructing the effective Hamiltonian using second-order many-body perturbation theory (CI+MBPT method)  \cite{DzuFlaKoz96}.

We separate the scalar dynamic polarizability $\alpha(\omega)$ into three parts:
\begin{equation}
\alpha(\omega) = \alpha_v(\omega) + \alpha_c(\omega) + \alpha_{vc}(\omega),
\label{alpha}
\end{equation}
where $\alpha_v$ is the valence polarizability, $\alpha_c$ is the ionic core polarizability, and a small term $\alpha_{vc}$
that corrects ionic core polarizability for the Pauli principle-violating excitations to occupied valence shells.

The valence part of the a.c. electric dipole polarizability of the $|0 \rangle$ state is
\begin{eqnarray}
\alpha_v (\omega) &=& 2\, \sum_k \frac { \left( E_k-E_0 \right)
|\langle \Phi_0 |D_0| \Phi_k \rangle|^2 }
      { \left( E_k-E_0 \right)^2 - \omega^2 }  \nonumber \\
 &=& \sum_k \left[ \frac {|\langle \Phi_0 |D_0| \Phi_k \rangle|^2 } {E_k - E_0 + \omega}
 +\frac {|\langle \Phi_0 |D_0| \Phi_k \rangle|^2 } {E_k - E_0 - \omega} \right]\!,
\label{Eqn_alpha}
\end{eqnarray}
where $D_0$ is the $z$-component of the effective electric dipole operator ${\bf D}$, defined in
atomic units ($\hbar=m=|e|=1$) as ${\bf D} = -{\bf r}$.
The effective (or ``dressed'') electric dipole operator
includes random-phase approximation (RPA), core-Brueckner ($\sigma$), structural radiation (SR),
and normalization corrections which are descibed in detail in ~\cite{DzuKozPor98}.
\begin{table*}
\caption{\label{Pol_stat} Contributions to the  $6s^2\,\ce{^1$S$_0}$, $5d6s\,\ce{^3$D$_1}$, and $5d6s\,\ce{^3$D$_2}$ scalar static  polarizabilities of Lu$^+$ in a.u.
The contributions to the valence polarizabilities of several lowest-lying intermediate states are listed separately with the corresponding absolute values of electric-dipole reduced matrix elements given in column labeled ``$D$''. The theoretical and experimental
\cite{RalKraRea11} transition energies are given in columns $\Delta E_{\rm th}$  and $\Delta E_{\rm expt}$.
The remaining valence contributions to the \ce{^1$S$_0} polarizability are given in row labeled ``Other''.
For the \ce{^3$D$_1} and \ce{^3$D$_2} polarizabilities we present the contribution of other (not explicitly listed in the table)
intermediate states with fixed total angular momentum $J_n$ in rows labeled ``Other ($J_n = 0,1,2,3$)''. In rows labeled
``Total ($J_n = 0,1,2,3$)'' we give total contribution of {\it all} intermediate states with fixed total angular momentum $J_n$.
In rows ``Total val.'' we present the total values of $\alpha_v$. The contributions from the $\alpha_{c}$ and
$\alpha_{vc}$ terms are listed together in rows labeled ``Core + Vc''.
The dominant contributions to the polarizabilities, listed in columns
$\alpha[\mathrm{A}]$ and $\alpha[\mathrm{B}]$, are calculated with the experimental~\cite{RalKraRea11}
and theoretical energies, respectively.}
\begin{ruledtabular}
\begin{tabular}{llccrdd}
\multicolumn{1}{l}{State}
&\multicolumn{1}{l}{Contribution} & \multicolumn{1}{c}{$\Delta E_{\rm th}$}
& \multicolumn{1}{c}{$\Delta E_{\rm expt}$} & \multicolumn{1}{c}{$D^{\rm a}$} &
\multicolumn{1}{c}{$\alpha[\mathrm{A}]$}
&\multicolumn{1}{c}{$\alpha[\mathrm{B}]$} \\
\hline \\ [-0.3pc]
$6s^2\,\ce{^1$S$_0}$ &$6s^2\,\ce{^1$S$_0} - 6s6p\,\ce{^3$P$^o_1}$& 29073 & 28503  & 0.820  &   3.45  &   3.39  \\
                &$6s^2\,\ce{^1$S$_0} - 6s6p\,\ce{^1$P$^o_1}$& 38862 & 38223  & 3.518  &  47.38  &  46.60   \\
                &$6s^2\,\ce{^1$S$_0} - 5d6p\,\ce{^3$D$^o_1}$& 46593 & 45532  & 0.811  &   2.11  &   2.07   \\
                &$6s^2\,\ce{^1$S$_0} - 5d6p\,\ce{^3$P$^o_1}$& 51285 & 50049  & 0.447  &   0.59  &   0.57   \\
                &$6s^2\,\ce{^1$S$_0} - 5d6p\,\ce{^1$P$^o_1}$& 60214 & 59122  & 1.354  &   4.54  &   4.46   \\
                & Other                           &       &        &        &   2.03  &   2.03    \\
                & Total val.                      &       &        &        &  60.10  &  59.11  \\
                & Core + Vc                       &       &        &        &   3.92  &   3.92   \\
                & Total                           &       &        &        &  64.02  &  63.03   \\
                & Recommended                     &       &        &        &         &  63.0    \\[0.5pc]

$5d6s\,\ce{^3$D$_1}$ &$5d6s\,\ce{^3$D$_1} - 6s6p\,\ce{^3$P$^o_0}$& 15297 & 15468  & 1.480  &   6.91  &   6.99  \\
                &$5d6s\,\ce{^3$D$_1} - 5d6p\,\ce{^3$P$^o_0}$& 38664 & 38167  & 1.892  &   4.57  &   4.52   \\
                & Other ($J_n = 0$)               &       &        &        &   0.33  &   0.33   \\
                & Total ($J_n = 0$)               &       &        &        &  11.82  &  11.83   \\[0.3pc]

                &$5d6s\,\ce{^3$D$_1} - 6s6p\,\ce{^3$P$^o_1}$& 16521 & 16707  & 1.287  &   4.83  &   4.89  \\
                &$5d6s\,\ce{^3$D$_1} - 5d6p\,\ce{^3$D$^o_1}$& 34041 & 33736  & 2.391  &   8.26  &   8.19  \\
                &$5d6s\,\ce{^3$D$_1} - 5d6p\,\ce{^3$P$^o_1}$& 38733 & 38253  & 2.089  &   5.56  &   5.49  \\
                & Other ($J_n = 1$)               &       &        &        &   0.59  &   0.59  \\
                & Total ($J_n = 1$)               &       &        &        &  19.25  &  19.16  \\[0.3pc]

                &$5d6s\,\ce{^3$D$_1} - 6s6p\,\ce{^3$P$^o_2}$& 20510 & 20657  & 0.351  &   0.29  &   0.29  \\
                &$5d6s\,\ce{^3$D$_1} - 5d6p\,\ce{^3$F$^o_2}$& 29925 & 29429  & 2.741  &  12.46  &  12.25  \\
                &$5d6s\,\ce{^3$D$_1} - 5d6p\,\ce{^1$D$^o_2}$& 34094 & 33662  & 1.716  &   4.27  &   4.21  \\
                &$5d6s\,\ce{^3$D$_1} - 5d6p\,\ce{^3$D$^o_2}$& 35488 & 35108  & 2.259  &   7.09  &   7.01  \\
                &$5d6s\,\ce{^3$D$_1} - 5d6p\,\ce{^3$P$^o_2}$& 39899 & 39405  & 0.555  &   0.38  &   0.38  \\
                & Other ($J_n = 2$)               &       &        &        &   4.50  &   4.50  \\
                & Total ($J_n = 2$)               &       &        &        &  28.99  &  28.65  \\[0.3pc]

                & Total val.                      &       &        &        &  60.05  &  59.64  \\
                & Core + Vc                       &       &        &        &   3.84  &   3.84  \\
                & Total                           &       &        &        &  63.89  &  63.48  \\
                & Recommended                     &       &        &        &         &  63.5   \\[0.5pc]

$5d6s\,\ce{^3$D$_2}$ &$5d6s\,\ce{^3$D$_2} - 6s6p\,\ce{^3$P$^o_1}$& 15867 & 16068  & 2.084  &   7.91  &   8.01  \\
                &$5d6s\,\ce{^3$D$_2} - 6s6p\,\ce{^1$P$^o_1}$& 25656 & 25788  & 0.814  &   0.75  &   0.76   \\
                &$5d6s\,\ce{^3$D$_2} - 5d6p\,\ce{^3$D$^o_1}$& 33387 & 33097  & 1.986  &   3.49  &   3.46   \\
                & Other ($J_n = 1$)               &       &        &        &   5.26  &   5.26   \\
                & Total ($J_n = 1$)               &       &        &        &  17.41  &  17.49   \\[0.3pc]

                &$5d6s\,\ce{^3$D$_2} - 6s6p\,\ce{^3$P$^o_2}$& 19857 & 20018  & 1.220  &   2.18  &   2.19  \\
                &$5d6s\,\ce{^3$D$_2} - 5d6p\,\ce{^3$F$^o_2}$& 29271 & 28790  & 2.552  &   6.62  &   6.51  \\
                &$5d6s\,\ce{^3$D$_2} - 5d6p\,\ce{^1$D$^o_2}$& 33440 & 33023  & 0.098  &   0.01  &   0.01  \\
                &$5d6s\,\ce{^3$D$_2} - 5d6p\,\ce{^3$D$^o_2}$& 34834 & 34469  & 2.653  &   5.97  &   5.91  \\
                & Other ($J_n = 2$)               &       &        &        &   3.71  &   3.71  \\
                & Total ($J_n = 2$)               &       &        &        &  18.49  &  18.34  \\[0.3pc]

                &$5d6s\,\ce{^3$D$_2} - 5d6p\,\ce{^3$F$^o_3}$& 33052 & 32483  & 3.727  &  12.52  &  12.30  \\
                &$5d6s\,\ce{^3$D$_2} - 5d6p\,\ce{^3$D$^o_3}$& 36720 & 36298  & 2.748  &   6.09  &   6.02  \\
                & Other ($J_n = 3$)               &       &        &        &   4.13  &   4.13  \\
                & Total ($J_n = 3$)               &       &        &        &  22.73  &  22.45  \\[0.3pc]

                & Total val.                      &       &        &        &  58.63  &  58.27  \\
                & Core + Vc                       &       &        &        &   3.84  &   3.84   \\
                & Total                           &       &        &        &  62.47  &  62.10   \\
                & Recommended                     &       &        &        &         &  62.1
\end{tabular}
$^{\rm a}$The values are obtained in the CI + all-order approximation and include
RPA, $\sigma$, SR, and normalization corrections.
\end{ruledtabular}
\end{table*}
In order to accurately  account for highly-excited  discrete states and a continuum we calculated $\alpha_v(\omega)$
using inhomogeneous equation in valence space rather than sum-over states formula given by Eq.~(\ref{Eqn_alpha})
We use the Sternheimer~\cite{Ste50} or Dalgarno-Lewis \cite{DalLew55}
method implemented in the framework of the CI+all-order approach following Ref.~\cite{KozPor99a}. Given the $\Phi_0$ wave function and energy
$E_0$ of the $|0 \rangle$ state, we find intermediate-state wave functions $\delta \psi_{\pm}$
from the inhomogeneous equation,
\begin{eqnarray}
|\delta \psi_{\pm} \rangle & = & \frac{1}{H_{\rm eff} - E_0 \pm \omega}\,
 \sum_k | \Phi_k \rangle \langle \Phi_k | D_0 | \Phi_0 \rangle \nonumber \\
&=&  \frac{1}{H_{\rm eff}- E_0 \pm \omega} \, D_0 | \Phi_0 \rangle .
\label{delpsi}
\end{eqnarray}
Using Eq.~(\ref{Eqn_alpha}) and $\delta \psi_{\pm}$ introduced
above, we obtain
\begin{equation}
\alpha_v (\omega ) = \langle \Phi_0 |D_0| \delta \psi_+ \rangle
+ \langle \Phi_0 |D_0| \delta \psi_- \rangle  \, ,
\label{alpha2}
\end{equation}
where superscript $v$ emphasizes that only excitations of the
valence electrons are included in the intermediate-state wave
functions $\delta \psi_{\pm}$ due to the presence of $H_{\rm eff}$.

\subsubsection{Static polarizabilities}
\label{statpolar}
In case of static polarizabilities, $\omega=0$, \eref{Eqn_alpha} is written as
\begin{eqnarray}
\alpha_v (0) &=& 2\, \sum_k \frac {|\langle \Phi_0 |D_0| \Phi_k \rangle|^2} {E_k-E_0}.
\label{stat_alpha}
\end{eqnarray}

While we do not use the sum-over-states approach in the calculation of the polarizabilities, it is important to establish the dominant contributions to the final values for the purpose of estimating theoretical uncertainties. We combine  the electric-dipole matrix elements and energies according to the sum-over-states formula,~\eref{stat_alpha}, for the valence polarizability to calculate the contributions of specific transitions between low-lying states and these are given in table~\ref{Pol_stat}. Remaining valence contributions of higher-lying states are given in rows labeled ``Other''.

We have carried out two calculations of the dominant contributions of the intermediate states to the polarizabilities. In the first calculation (Column $\alpha$[B] in Table~\ref{Pol_stat}) we used our theoretical values of the energy levels in the denominator of~\eref{stat_alpha}. In the second calculation (Column $\alpha$[A] in Table~\ref{Pol_stat}) we used experimental energies, where available. Corresponding  theoretical and experimental~\cite{RalKraRea11} transition energies are given in columns $\Delta E_{\rm th}$ and $\Delta E_{\rm expt}$ in cm$^{-1}$. The difference between the results is -1.6\% for the \ce{^1$S$_0} polarizability
and -0.6\% for the \ce{^3$D$_1} and \ce{^3$D$_2} polarizabilities, demonstrating that deviation of our theoretical energies from the experimental values does not significantly affect overall accuracy of the polarizabilities.
The absolute values of the corresponding reduced electric-dipole matrix elements in a.u. are listed in
columns labeled ``$D$''.  These are calculated using the CI + all-order method and include RPA, $\sigma$, SR, and normalization corrections. Calculation of the RPA, $\sigma$, and SR corrections is discussed in~\cite{DzuKozPor98}.


The contributions from $\alpha_{c}$ and $\alpha_{vc}$ terms evaluated in the RPA approximation are listed together in rows labeled ``Core +Vc''.
Taking into account that the main contribution to the \ce{^3$D$_1} and \ce{^3$D$_2} levels comes from
the $5d_{3/2} 6s$ configuration (99\% and 80\%, respectively), we determined $\alpha_{vc}$ terms for
the ${^3D_{1,2}}$ polarizabilities as $\alpha_{vc}(5d_{3/2})+\alpha_{vc}(6s)$.
In rows labeled ``Total'' we present the total values of the scalar static \ce{^1$S$_0}, \ce{^3$D$_1}, and \ce{^3$D$_2} polarizabilities. Our final values are given in rows labeled ``Recommended''.

To determine uncertainties of the polarizabilities we have also calculated them using two other approximations:
the CI+MBPT+RPA and CI+all-order+RPA. In both cases only RPA corrections were included. CI+MBPT method omits
higher-order core-valence correlations.
The results obtained in the
CI+MBPT+RPA, CI+all-order+RPA, and CI+all-order+AC approximations (where abbreviation ``AC'' means {\it all corrections}, including RPA, $\sigma$, SR, and normalization) are presented in~\tref{Polariz} in columns (1), (2), and (3), correspondingly.

We consider the results  obtained in the CI+all-order+AC approximation as the final values according to Sr study \cite{SafPorSaf13}.
Comparison of the data in columns (2) and (3) in~\tref{Polariz} illustrates that the corrections beyond RPA only slightly
change the values of the \ce{^1$S$_0} and ${^3D_{1,2}}$ polarizabilities.
We estimate the polarizability  uncertainties as the spread of the results in columns (1), (2), and (3).

\begin{table}
\caption{\label{Polariz} The scalar ($\alpha_0$) and tensor ($\alpha_2$) polarizabilities,
obtained in the CI+MBPT+RPA, CI+all-order+RPA, and CI+all-order+AC approximations (where ``AC'' means {\it all corrections})
are presented (in a.u.) in columns (1), (2), and (3), correspondingly. Final (recommended) values are given in the last column.
The uncertainties are given in parentheses.}
\begin{ruledtabular}
\begin{tabular}{lcccd}
\multicolumn{1}{l}{Polarizability} & \multicolumn{1}{c}{(1)} & \multicolumn{1}{c}{(2)} &
\multicolumn{1}{c}{(3)} & \multicolumn{1}{c}{Final} \\
\hline \\,[-0.3pc]
$\alpha_0(6s^2\,\ce{^1$S$_0})$             & 62.5 &  63.3  &  63.0  &  63.0(0.8)  \\[0.5pc]

$\alpha_0(5d6s\,\ce{^3$D$_1})$             & 61.5 &  64.3  &  63.5  &  63.5(2.8)  \\
$\alpha_2(5d6s\,\ce{^3$D$_1})$             & -4.8 &  -5.2  &  -5.1  & -5.1(4) \\[0.5pc]

$\alpha_0(5d6s\,\ce{^3$D$_2})$             & 60.3 &  62.9  &  62.1  &  62.1(2.6)  \\
$\alpha_2(5d6s\,\ce{^3$D$_2})$             & -5.1 &  -5.7  &  -5.6  & -5.6(6) \\[0.5pc]

$\alpha_0(^3\!D_1)-\alpha_0(\ce{^1$S$_0})$ & -1.0 &   1.0  &   0.5  &  0.5  \\
$\alpha_0(\ce{^3$D$_2})-\alpha_0(\ce{^1$S$_0})$ & -2.2 &  -0.4  &  -0.9  & -0.9
\end{tabular}
\end{ruledtabular}
\end{table}

\subsubsection{Dynamic polarizabilities}
\label{dynpol}
We have also calculated the dynamic scalar and tensor polarizabilities for the \ce{^1$S$_0}, \ce{^3$D$_1}, and \ce{^3$D$_2}
states for the  wavelengths of experimental interest. The results, presented in~\tref{Pol_dyn}, are obtained in the
framework of the CI+all-order+AC approximation, i.e., all corrections to the matrix elements are included.
\begin{table}
\caption{\label{Pol_dyn} The dynamic scalar ($\alpha_0$), tensor ($\alpha_2$), and differential
$\Delta_{1,2} \equiv \alpha_0(\ce{^3D_{1,2}}) - \alpha_0(\ce{^1$S$_0}) $ polarizabilities (in a.u.), obtained
in the CI+all-order+AC approximation, are calculated for the wavelengths (frequencies) given in 1st (2nd)
line.}
\begin{ruledtabular}
\begin{tabular}{llddddd}
 $\lambda$ (nm)  &             &\multicolumn{1}{c}{847.7}
                                       &\multicolumn{1}{c}{1064}
                                                &\multicolumn{1}{c}{1560}
                                                         &\multicolumn{1}{c}{1760}
                                                                  &\multicolumn{1}{c}{10600} \\
 $\omega$ (a.u.) &             &\multicolumn{1}{c}{0.05375}
                                       &\multicolumn{1}{l}{0.04282}
                                                &\multicolumn{1}{l}{0.02921}
                                                         &\multicolumn{1}{c}{0.02589}
                                                                  &\multicolumn{1}{c}{0.00430}\\
\hline \\ [-0.3pc]
 $6s^2\,\ce{^1$S$_0}$ & $\alpha_0$  &  68.9  &  66.6 &  64.6  &  64.3  & 63.0 \\[0.5pc]

 $5d6s\,\ce{^3$D$_1}$ & $\alpha_0$  &  85.4  &  73.9 &  67.6  &  66.6  & 63.6 \\[0.1pc]
                 & $\alpha_2$  & -13.0  &  -8.3 &  -6.2  &  -5.9  & -5.1 \\[0.5pc]

 $5d6s\,\ce{^3$D$_2}$ & $\alpha_0$  &  79.6  &  70.9 &  65.6  &  64.8  & 62.2 \\[0.1pc]
                 & $\alpha_2$  & -14.1  &  -9.1 &  -6.8  &  -6.5  & -5.6 \\[0.5pc]

                 &$\Delta_1$   &  16.5  &   7.3 &   2.9  &   2.3  &  0.5 \\
                 &$\Delta_2$   &  10.7  &   4.3 &   1.0  &   0.5  & -0.9
\end{tabular}
\end{ruledtabular}
\end{table}

\begin{figure}[!htb]
\includegraphics[width=\linewidth]{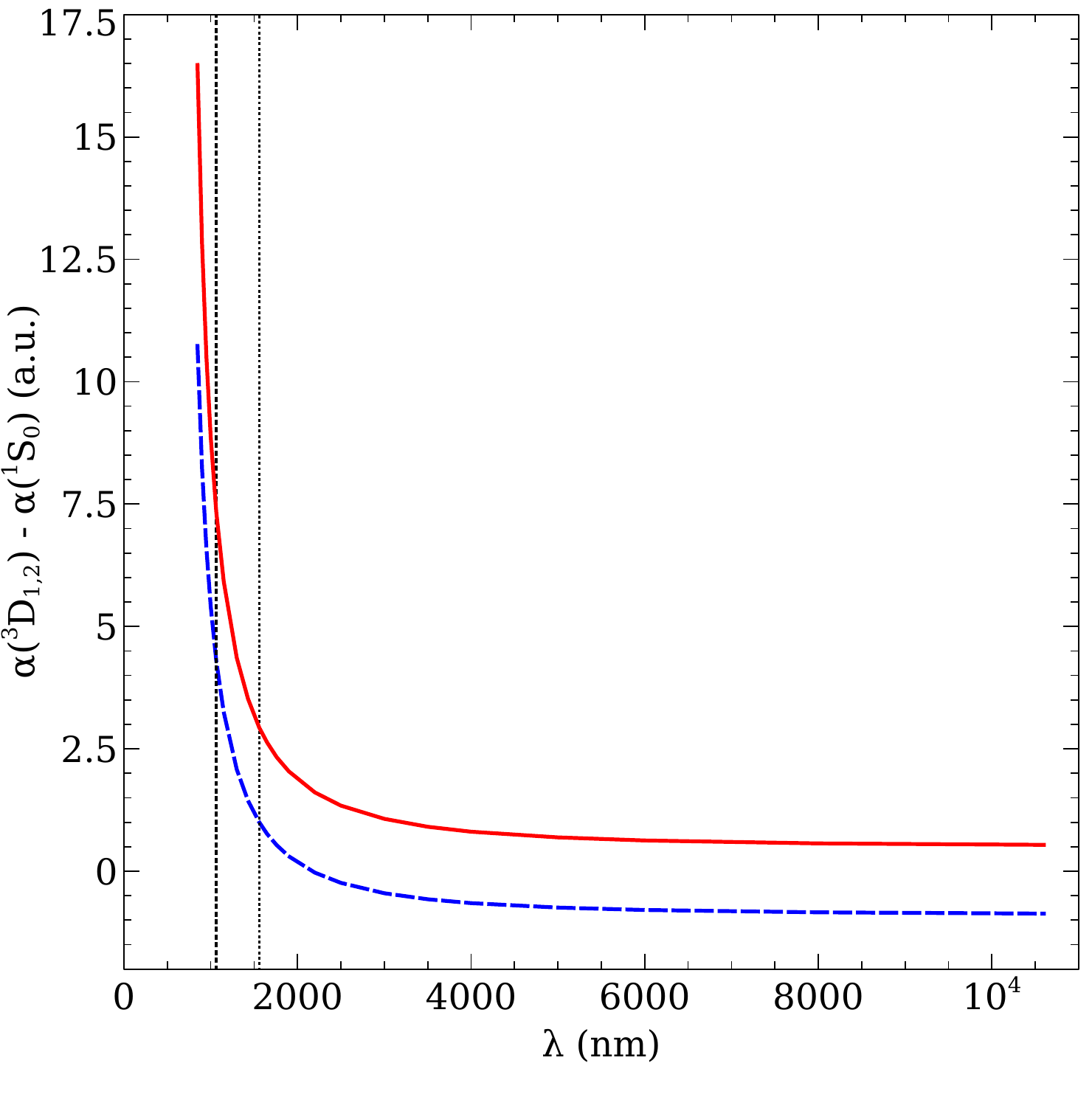}
\caption{(Color online) Differential scalar polarizabilities $\alpha(\ce{^3$D$_1})-\alpha(\ce{^1$S$_0})$ and $\alpha(\ce{^3$D$_2})-\alpha(\ce{^1$S$_0})$
represented by red solid and blue dashed lines, respectively, vs. the wavelength $\lambda$.
The vertical dotted lines correspond to $\lambda$ = 1064 and 1560 nm.}
\label{fig:3D-1S0}
\end{figure}

In Fig.~\ref{fig:3D-1S0} we plot differential scalar polarizabilities $\alpha(\ce{^3$D$_1})-\alpha(\ce{^1$S$_0})$ and $\alpha(\ce{^3$D$_2})-\alpha(\ce{^1$S$_0})$
represented by red solid and blue dashed lines, respectively, vs. the wavelength $\lambda$. The vertical dotted lines correspond
to $\lambda$ = 1064 and 1560 nm.
\subsection{Lifetimes of the low-lying states}
\label{Sect_lifetimes}
In Table~\ref{branching}, we list the lifetimes $\tau$ of the low-lying $6s6p\,{ ^{3,1}P^o_J}$ and $5d6s\,{^{3,1}D_J}$
states together with most important reduced matrix elements, and relevant transition rates and branching ratios.

The $E1$, $E2$, and $M1$ transition probabilities (in s$^{-1}$) are obtained in terms of reduced matrix elements (MEs)
of the electric-dipole, electric-quadrupole, and magnetic-dipole operators, and transition frequencies $\omega$ as
\begin{eqnarray*}
W_{E1} (\gamma J \rightarrow \gamma' J') = 2.02613 \cdot 10^{-6} \,
\frac{\omega^3 \, \langle \gamma' J' ||D||\gamma J \rangle^2}{2J+1},
\end{eqnarray*}%
\begin{eqnarray*}
W_{E2} (\gamma J \rightarrow \gamma' J') = 1.11995 \cdot 10^{-22} \,
\frac{\omega^5 \, \langle \gamma' J' ||Q_E|| \gamma J \rangle^2}{2J+1},
\end{eqnarray*}%
\begin{eqnarray*}
W_{M1} (\gamma J \rightarrow \gamma' J') = 2.69735 \cdot 10^{-11} \,
\frac{\omega^3 \, \langle \gamma' J' ||\mu|| \gamma J \rangle^2}{2J+1}.
\end{eqnarray*}%
In these equations, MEs of the $E1$ and $E2$ operators are expressed in a.u., MEs of the $M1$ operator in Bohr magnetons ($\mu_0$), and the decay rates $\omega$ are expressed in cm$^{-1}$.

We determine the lifetimes, listed in the last column of Table~\ref{branching}, as $\tau = 1/W_{\rm tot}$ with
$W_{\rm tot} \equiv \sum_k W_k$, where $W_k$ are the individual decay rates. The branching ratios, $B_k$, are determined as $B_k = W_k/W_{\rm tot}$.

\begin{table*}[!htb]
\caption{\label{branching}  The energies (in cm$^{-1}$) are counted from the ground $6s^2\,{ ^1S_0}$ state.
5th column gives type of transition. The reduced MEs of $E1$ and $E2$ operators (in a.u.) and
$M1$ operator (in $\mu_0$) are presented in 6th column. The individual decay rates, $W_k$ (in s$^{-1}$),
branching ratios ($B_k$), and lifetimes ($\tau$) are listed in columns 7-9.
These quantities are evaluated in the CI+all-order+AC approximation. The numbers in brackets represent powers of 10.}
\begin{ruledtabular}
\begin{tabular}{rcrcccccc}
\multicolumn{2}{c}{Upper level} & \multicolumn{2}{c}{Lower level} & \multicolumn{1}{c}{Trans-n} & \multicolumn{1}{c}{ME}
& \multicolumn{1}{c}{$W_k$} & \multicolumn{1}{c}{$B_k$} & \multicolumn{1}{c}{$\tau$} \\
\multicolumn{1}{r}{Term} & \multicolumn{1}{c}{Energy} & \multicolumn{1}{r}{Term} & \multicolumn{1}{c}{Energy} & \multicolumn{1}{c}{}
& \multicolumn{1}{c}{} & \multicolumn{1}{c}{s$^{-1}$} & \multicolumn{1}{c}{} & \multicolumn{1}{c}{} \\
\hline \\ [-0.5pc]
\multicolumn{9}{c}{Transitions from the even-parity states} \\[0.4pc]
$6s5d\,\ce{^3$D$_1}$   & 11796 & $6s^2\,\ce{^1$S$_0}$  &   0   & $M1$ & 0.0006& 5.14[-6]  &  1.00 &  1.95[+5] (s) \\[0.4pc]

$6s5d\,\ce{^3$D$_2}$   & 12435 & $6s^2\,\ce{^1$S$_0}$  &   0   & $E2$ & 2.509 & 4.19[-2]  &  0.88 &  20.9 (s) \\
                  &       & $6s5d\,\ce{^3$D$_1}$  & 11796 & $E2$ & 4.523 & 4.88[-8]  &  0.00 &           \\
                  &       &                  & 11796 & $M1$ & 2.055 & 5.94[-3]  &  0.12 &           \\[0.4pc]

$6s5d\,\ce{^3$D$_3}$   & 14199 & $6s5d\,\ce{^3$D$_1}$  & 11796 & $E2$ & 1.601 & 3.29[-6]  &  0.00 &  10.8 (s) \\
                  &       & $6s5d\,\ce{^3$D$_2}$  & 12435 & $E2$ & 4.969 & 6.74[-6]  &  0.00 &           \\
                  &       &                  & 12435 & $M1$ & 2.094 & 9.27[-2]  &  1.00 &           \\[0.4pc]

$6s5d\,\ce{^1$D$_2}$   & 17333 & $6s^2\,\ce{^1$S$_0}$  &   0   & $E2$ & 10.63 & 3.96      &  0.92 &  0.23 (s) \\
                  &       & $6s5d\,\ce{^3$D$_1}$  & 11796 & $E2$ & 1.018 & 1.21[-4]  &  0.00 &           \\
                  &       &                  & 11796 & $M1$ & 0.524 & 2.51[-1]  &  0.06 &           \\
                  &       & $6s5d\,\ce{^3$D$_2}$  & 12435 & $E2$ & 1.319 & 1.10[-4]  &  0.00 &           \\
                  &       &                  & 12435 & $M1$ & 0.218 & 3.01[-2]  &  0.01 &           \\
                  &       & $6s5d\,\ce{^3$D$_3}$  & 14199 & $E2$ & 1.327 & 1.19[-5]  &  0.00 &           \\
                  &       &                  & 14199 & $M1$ & 0.531 & 4.68[-2]  &  0.01 &           \\[0.4pc]

\multicolumn{9}{c}{Transitions from the odd-parity states} \\[0.4pc]

$6s6p\,\ce{^3$P$^o_0}$ & 27264 &  $5d6s\,\ce{^3$D$_1}$ & 11796 & $E1$ & 1.480 & 1.64[+7]  &  1.00 &  61.0 (ns) \\[0.4pc]

$6s6p\,\ce{^3$P$^o_1}$ & 28503 &  $6s^2\,\ce{^1$S$_0}$ &     0 & $E1$ & 0.820 & 1.05[+7]  &  0.38 &  35.7 (ns) \\
                  &       &  $5d6s\,\ce{^3$D$_1}$ & 11796 & $E1$ & 1.287 & 5.22[+6]  &  0.19 &  \\
                  &       &  $5d6s\,\ce{^3$D$_2}$ & 12435 & $E1$ & 2.084 & 1.22[+7]  &  0.43 &   \\
                  &       &  $5d6s\,{^1D_2}$ & 17333 & $E1$ & 0.329 & 1.00[+5]  &  $<0.01$ &  \\[0.4pc]

$6s6p\,\ce{^3$P$^o_2}$ & 32453 &  $5d6s\,\ce{^3$D$_1}$ & 11796 & $E1$ & 0.351 & 4.40[+5]  &  0.02 &  35.8 (ns) \\
                  &       &  $5d6s\,\ce{^3$D$_2}$ & 12435 & $E1$ & 1.220 & 4.84[+6]  &  0.17 &  \\
                  &       &  $5d6s\,\ce{^3$D$_3}$ & 14199 & $E1$ & 3.015 & 2.24[+7]  &  0.80 &  \\
                  &       &  $5d6s\,{^1D_2}$ & 17333 & $E1$ & 0.445 & 2.77[+5]  &  0.01 &  \\ [0.4pc]

$6s6p\,\ce{^1$P$^o_1}$ & 38224 &  $6s^2\,{ ^1S_0}$ &     0 & $E1$ & 3.518 & 4.67[+8]  &    0.95&  20.8 (ns) \\
                  &       &  $5d6s\,\ce{^3$D$_2}$ & 12435 & $E1$ & 0.814 & 7.67[+6]  &    0.02&  \\
                  &       &  $5d6s\,{^1D_2}$ & 17333 & $E1$ & 0.994 & 6.08[+6]  &    0.01&  
\end{tabular}
\end{ruledtabular}
\end{table*}

To estimate uncertainty of theoretical values, we calculate the following decay rates:
\begin{eqnarray}
W_0 &\equiv& W(6s6p\,\ce{^3$P$^o_1} \rightarrow 6s^2\,\ce{^1$S$_0}), \nonumber \\
W_1 &\equiv& W(6s6p\,\ce{^3$P$^o_1} \rightarrow 5d6s\,\ce{^3$D$_1}), \nonumber \\
W_2 &\equiv& W(6s6p\,\ce{^3$P$^o_1} \rightarrow 5d6s\,\ce{^3$D$_2}), \nonumber \\
W_3 &\equiv& W(6s6p\,\ce{^3$P$^o_1} \rightarrow 5d6s\,\ce{^1$D$_2}),
\label{Wk}
\end{eqnarray}
using three different methods: (1) CI+MBPT+RPA, (2) CI+all-order+RPA, and (3) CI+all-order+AC approximations. These results are given in~\tref{BR}. The spread of the values calculated in these approximations (1-3) gives an estimate of the uncertainties in the final results. Comparing the results obtained in the CI+all-order+RPA (2) and CI+all-order+AC (3)
approximations, we find that the corrections beyond RPA play a very insignificant role and we take the results presented in columns labeled (3) as final.

\begin{table*}[ht]
\caption{\label{BR} The  decay rates $W_k$ determined by~\eref{Wk} (in s$^{-1}$)
and branching ratios, obtained in (1) CI+MBPT+RPA, (2) CI+all-order+RPA, and (3) CI+all-order+AC
approximations, are listed. The experimental values from section~\ref{3P1Measurements} are given in the last column. The numbers in brackets represent
powers of 10.}
\begin{ruledtabular}
\begin{tabular}{ccccccccc}
\multicolumn{4}{c}{\multirow{1}{*}{Probabilities}} &
\multicolumn{5}{c}{\multirow{1}{*}{Branching ratios}} \\
\cline{1-4} \cline{5-9} \\ [-0.6pc]
 &\multicolumn{1}{c}{$(1)$} &\multicolumn{1}{c}{$(2)$} &\multicolumn{1}{c}{$(3)$}
&& \multicolumn{1}{c}{$(1)$} &\multicolumn{1}{c}{$(2)$} &\multicolumn{1}{c}{$(3)$} &\multicolumn{1}{c}{Experim.}\\
\hline \\ [-0.3pc]
 $W_0$ &  1.24[+7] & 1.07[+7] & 1.05[+7]  & $B_0$ & 0.408   & 0.375   & 0.376   & 0.392 \\
 $W_1$ &  5.33[+6] & 5.33[+6] & 5.22[+6]  & $B_1$ & 0.176   & 0.186   & 0.186   & 0.186 \\
 $W_2$ &  1.25[+7] & 1.25[+7] & 1.22[+7]  & $B_2$ & 0.413   & 0.435   & 0.434   & 0.418 \\
 $W_3$ &  9.59[+4] & 1.02[+5] & 1.00[+5]  & $B_3$ & 0.00316 & 0.00357 & 0.00357 & 0.00436
\end{tabular}
\end{ruledtabular}
\end{table*}

\subsection{Hyperfine quenching of a state with $J=0$}
\label{Quench}
Hyperfine quenching rate of a state with total angular momentum $J=0$ is given by
\begin{eqnarray}
W( \gamma J=0 \rightarrow \gamma' J') &=&
\frac{4 \alpha^3 \omega^3}{3} \, \frac{1}{(2J+1)(2I+1)} \nonumber \\
&\times& \sum_k \frac{\langle I||N^{(k)}||I\rangle^2}{3(2k+1)} \left\vert S_{k}\right\vert^2 ,
\label{Wq}
\end{eqnarray}%
where $\omega$ is the $(\gamma J=0 \rightarrow \gamma' J')$ transition frequency and
\begin{eqnarray}
&&S_{k} \equiv \sqrt{\frac{3}{2k+1}}\sum_{\gamma _{n}}\frac{\langle \gamma
^{\prime }J^{\prime }||D||\gamma_{n}J_{n} \rangle \langle \gamma
_{n}J_{n}||T^{\left( k\right) }||\gamma J=0\rangle }{E_{n}-E_{\gamma J}} \nonumber \\
&&+\underset{\gamma _{m}J_{m}\neq \gamma ^{\prime }J^{\prime }}{\sum }\frac{%
\langle \gamma ^{\prime }J^{\prime }||T^{\left( k\right) }||\gamma
_{_{m}}J_{_{m}}\rangle \langle \gamma _{_{m}}J_{_{m}}||D||\gamma
J=0\rangle }{E_{_{m}}-E_{\gamma ^{\prime }J^{\prime }}},
\label{Sk}
\end{eqnarray}
where $Q$, $\mu$, and $T^{(k)}$ are defined in Appendix~\ref{Te}.  The $^{175}$Lu$^+$ ion has the nuclear spin $I=7/2$. Its nuclear magnetic moment $\mu$, expressed in
nuclear magnetons $\mu_N$, is $\mu/\mu_N = 2.2323(11)$~\cite{NuclearDipole} and the nuclear quadrupole moment
$Q = 3.49(2)$ barn~\cite{NuclearQuad}.

In~\tref{MEs} we list absolute values of the reduced matrix elements of $ {T^{(1)}}$ and $T^{(2)}$ operators.
To illustrate the role of different corrections, we carried out 3 calculations and found
the MEs in (1) CI+MBPT+RPA, (2) CI+all-order+RPA, and (3) CI+all-order+AC approximations.
Respective values are listed in the table in columns labeled ``(1)'', ``(2)'', and ``(3)''.

In contrast with the MEs of the electric-dipole operator, the corrections beyond RPA ($\sigma$, SR, and normalization) are large for the matrix elements of $T^{(1)}$ and $T^{(2)}$ operators.
They contribute to large MEs at the level of 10\% and even more
to smaller MEs. In particular, it was essential to account for the structural radiation (SR) corrections for calculating the MEs of the $T^{(2)}$ operator between $D_J$ states. The SR contributions  are
$\sim$~20\% to $\langle 5d6s\,\ce{^3$D$_1} ||T^{(2)}||5d6s\,{^3D_{2,3}} \rangle$ and
40\% to $\langle 5d6s\,\ce{^3$D$_1} ||T^{(2)}||5d6s\,{^1D_{2}} \rangle$.

We note that while the RPA corrections were calculated to all orders,  the corrections beyond RPA were obtained only
in the 2nd order of MBPT, which usually overestimates respective contribution. For this reason our final (recommended)
values are based on the results obtained in the CI+all-order+RPA approximation while the assigned uncertainties
are determined as the differences between the CI+all-order+RPA and CI+all-order+AC values.
\begin{table}[ht]
\caption{\label{MEs} The absolute values of MEs of the $ {T^{(1)}}$ and $T^{(2)}$ operators (in MHz) obtained in
(1) CI+MBPT+RPA, (2) CI+all-order+RPA, and (3) CI+all-order+AC approximations are presented in columns labeled ``(1)'',
''(2)'', and ``(3)'', correspondingly. The final values and assigned uncertainties (in parentheses) are given
in last column.}
\begin{ruledtabular}
\begin{tabular}{lcccc}
\multicolumn{1}{c}{ME}
                                                               &  (1)  &  (2)   & (3)    & Final  \\
\hline \\ [-0.3pc]
$\langle 6s6p\,\ce{^3$P$^o_1} ||\,{T^{(1)}}|| 6s6p\,\ce{^3$P$^o_0} \rangle$  & 12427 & 11964  & 10833  & 11960(1000)  \\[0.3pc]

$\langle 6s6p\,\ce{^1$P$^o_1} ||\,{T^{(1)}}|| 6s6p\,\ce{^3$P$^o_0} \rangle$  &  4324 &  4352  &  3965  &  4350(400)   \\[0.3pc]

$\langle 5d6s\,\ce{^3$D$_1} ||\,{T^{(1)}}|| 5d6s\,\ce{^3$D$_2} \rangle$      & 19465 & 18682  & 16780  & 18680(1900) \\[0.3pc]

$\langle 6s6p\,\ce{^3$P$^o_2} ||\,{T^{(2)}}|| 6s6p\,\ce{^3$P$^o_0} \rangle$  &  1789 &  1780  &  1672  & 1780(110)   \\[0.3pc]

$\langle 5d6s\,\ce{^3$D$_1} ||\,{T^{(2)}}|| 5d6s\,\ce{^3$D$_3} \rangle$      &  200  &  198   &  250   &  200(50)   \\[0.3pc]

$\langle 5d6s\,\ce{^3$D$_1} ||\,{T^{(1)}}|| 5d6s\,\ce{^1$D$_2} \rangle$      & 10718 & 10618  &  9750  & 10620(870) \\[0.3pc]

$\langle 5d6s\,\ce{^3$D$_1} ||\,{T^{(2)}}|| 5d6s\,\ce{^1$D$_2} \rangle$      &   76  &   70   &   116  &   70(45)
\end{tabular}
\end{ruledtabular}
\end{table}

We determined the hyperfine quenching rates for the $6s6p\,\ce{^3$P$^o_0}$ state. We present the results obtained
in (1) CI+MBPT+RPA, (2) CI+all-order+RPA, and (3) CI+all-order+AC approximations in~\tref{Qr}. The probability of the main
$E1$ ($6s6p$\,\ce{^3$P$^o_0} - $5d6s$\,\ce{^3$D$_1}) transition, $W^{(0)}$,  is given in the 1st line. The quenching rates of the
$6s6p$\,\ce{^3$P$^o_0} - $6s^2$\,\ce{^1$S$_0}, $5d6s$\,\ce{^3$D$_{2,3}}, $5d6s$\,\ce{^1$D$_2} transitions, calculated using Eqs.~(\ref{Wq}) and (\ref{Sk}), are listed in lines 2-5, correspondingly. We sum all listed hyperfine quenching rates and present
in last line of the table the branching ratio of this sum to $W^{(0)}$.

As we  discussed above we consider the results obtained at the CI+all-order+RPA stage as the final (recommended) values as the calculations of other corrections beyond RPA is unreliable for the matrix elements of of the $T^{(k)}$ operators.
The uncertainties were estimated as the largest difference between the CI+all-order+RPA results
and the CI+MBPT+RPA and CI+all-order+AC values.
\begin{table}[htb]
\caption{\label{Qr} The $6s6p\,\ce{^3$P$^o_0} - 5d6s\,\ce{^3$D$_1}$ transition probability, $W^{(0)}$, is given in first line
(in s$^{-1}$). The quenching rates of the $6s6p\,\ce{^3$P$^o_0} - 6s^2\,\ce{^1$S$_0},\, 5d6s\,\ce{^3$D$_{2,3}}, \, 5d6s\,\ce{^1$D$_2}$ transitions,
obtained in (1) CI+MBPT+RPA, (2) CI+all-order+RPA, and (3) CI+all-order+AC approximations, are listed (in s$^{-1}$) on lines 2-5, correspondingly. The branching ratios (BR) of the hyperfine quenching rates to $W^{(0)}$ are presented in last line.  The uncertainties are given in parentheses. The numbers in brackets represent powers of 10.}
\begin{ruledtabular}
\begin{tabular}{lllll}
&\multicolumn{1}{c}{\multirow{1}{*}{(1)}} & \multicolumn{1}{c}{\multirow{1}{*}{(2)}}
& \multicolumn{1}{c}{\multirow{1}{*}{(3)}} & \multicolumn{1}{c}{\multirow{1}{*}{Final}} \\
\hline \\ [-0.3pc]
$\,\ce{^3$P$^o_0} - \,\ce{^3$D$_1}$ & 1.69[+7] & 1.68[+7]  &  1.64[+7] & 1.68(4)[+7] \\[0.3pc]
$\,\ce{^3$P$^o_0} - \,\ce{^1$S$_0}$ & 3.55     & 2.96      &  2.38     & 2.96(59) \\[0.1pc]
$\,\ce{^3$P$^o_0} - \,\ce{^3$D$_2}$ & 3.26     & 2.96      &  2.31     & 2.96(65) \\[0.1pc]
$\,\ce{^3$P$^o_0} - \,\ce{^3$D$_3}$ & 0.0011   & 0.0010    &  0.0008   & 0.0010(2) \\[0.1pc]
$\,\ce{^3$P$^o_0} - \,\ce{^1$D$_2}$ & 0.051    & 0.050     &  0.041    & 0.050(9) \\[0.3pc]
BR                      & 4.05[-7] & 3.55[-7]  &  2.88[-7] & 3.55(65)[-7]
\end{tabular}
\end{ruledtabular}
\end{table}

\section{Discussion}
We have measured several key properties of ${^{175}\mathrm{Lu}^+}$ that are relevant to practical clock operation with this ion.  Hyperfine induced mixing results in a small decay rate out the detection channel which is dominated by decay into the \ce{^1$S$_0} and \ce{^3$D$_2} levels.  This rate provides a fundamental limit to the detection error rate given by $w/(W q)$ which is the probability the ion is pumped dark with zero photons detected.  For our current collection efficiency this gives a limit of $1.0\times 10^{-4}$.  The measured decay rate does not impose any limitations on cooling as most decays are to  \ce{^1$S$_0} and \ce{^3$D$_2} which can be quickly repumped.  

During clock operation occupation of the \ce{^1$D$_2} level would mostly occur due to repumping to \ce{^3$D$_1}.  This occurs with a probability of approximately $2\%$.  In typical clock operation, occupation is split between ground and excited state.  So, on average, $1\%$ of the cycles will be compromised provided the cycle time is large enough for the ion to decay from \ce{^1$D$_2} with reasonable probability before the next cycle begins.  Occupation of the \ce{^3$D$_3} would result in significant dead-time.  However, based on the analysis here, these events are infrequent, happening only once every $10^4$ clock cycles.  In a multi-ion clock \cite{MDB2} these considerations would only result in a very small number fluctuations such that additional repump lasers would not be necessary.

Calculation of polarizabilities given in section~\ref{Polar} indicate that the differential scalar polarizability, $\Delta \alpha$, for the \ce{^1$S$_0} to \ce{^3$D$_1} may not be as reported in \cite{Dzuba} and this would have immediate consequences for the proposal given in \cite{MDB2}.  It is therefore essential to obtain an experimental value for this quantity.  We have given calculations of $\Delta \alpha$ at a number of wavelengths that are readily accessible to us.  Measurement of $\Delta \alpha$ at these wavelengths would serve as a useful benchmark for the calculations given here.  

Intuitively we can expect $\Delta \alpha$ to be more negative for the \ce{^1$S$_0} to \ce{^3$D$_2} transition and we have also given associated calculations for this case as well.  Measurements and calculations here demonstrate a suitable lifetime for clock operation.  Although this transition would be more technically difficult to implement, systematic shifts would be significantly lower than for the \ce{^1$S$_0} to \ce{^3$D$_2} case.  Contributions from the \ce{^3$D$_1} and \ce{^3$D$_3} levels have opposite sign resulting in a partial cancellation of the residual magnetic field shift for the average frequency.  Furthermore, due to the reduced lifetime relative to the \ce{^3$D$_1} level, much less intensity is needed to drive the \ce{^1$S$_0} to \ce{^3$D$_2} transition resulting in a substantial reduction in the AC stark shift from the probe beam itself.
\begin{acknowledgements}
We acknowledge the support of this work by the National Research Foundation and the Ministry of Education of Singapore. This work was supported in part by U.S. NSF grant PHY-1520993 and the Australian Research Council.
\end{acknowledgements}
\appendix
\section{The Hyperfine Interaction}
\label{Te}
The hyperfine structure (HFS) coupling due to nuclear multipole moments may be represented as a scalar product of
two tensors of rank $k$,
\begin{equation*}
H_{\rm hfs} = \sum_k H_{{\rm hfs},k} =
\sum_k \left( \mathbf{N}^{(k)}\cdot \mathbf{T}^{\left( k\right) }\right) ,
\end{equation*}
where $\mathbf{N}^{(k)}$ and $\mathbf{T}^{\left( k\right) }$ act in the space
of nuclear and electronic coordinates, respectively. Using this expression we
 write the $H_{\mathrm{hfs}}$ matrix element as
\begin{eqnarray*}
&&\langle \gamma' IJ';FM_F |H_{\mathrm{hfs}}|\gamma IJ;FM_F \rangle = (-1)^{I+J'+F} \\
&\times& \sum_k \langle I ||N^{(k)}|| I \rangle \langle \gamma' J' ||T^{(k)}||\gamma J\rangle
\left\{
\begin{tabular}{lll}
$I$ & $I$  & $k$ \\
$J$ & $J'$ & $F$
\end{tabular}
\right\} ,
\end{eqnarray*}
where ${\bf I}$ is the nuclear spin, ${\bf J}$ is the total angular momentum of the electrons, ${\bf F} = {\bf J} + {\bf I}$,
$M_F$ is the projection of the total momentum ${\bf F}$ to quantization axis, and
$\gamma$ encapsulates all other atomic quantum numbers.

Below, we restrict the treatment of  $H_{\mathrm{hfs}}$ to the first two terms in the sum over
$k$, i.e., we consider only the interaction of magnetic dipole and
electric quadrupole nuclear moments with the electrons. Thus,
\begin{equation*}
H_{\mathrm{hfs}}\approx \mathbf{N}^{(1)}\cdot \mathbf{T}^{\left( 1\right) }+%
\mathbf{N}^{(2)}\cdot \mathbf{T}^{\left( 2\right) } .
\end{equation*}
It is convenient to express the matrix elements $\langle I ||N^{(1)}||I\rangle $
and $\langle I ||N^{(2)}|| I\rangle $ through the
nuclear magnetic dipole moment $\boldsymbol \mu$ and nuclear electric quadrupole moment
$Q$, respectively. They are defined as follows
\begin{eqnarray*}
\mu &=& \langle IM_I=I|{\boldsymbol \mu}_z|IM_I=I \rangle
=\left(
\begin{tabular}{rll}
$I$  & $1$ & $I$ \\
$-I$ & $0$ & $I$%
\end{tabular}
\right) \langle I ||\mu|| I \rangle \nonumber \\
&=& \sqrt{\frac{I}{(2I+1)(I+1)}} \, \langle I ||\mu||I \rangle , \\
Q &=& 2 \langle IM_I=I |Q_0^{(2)}| IM_I=I \rangle =
2 \left(
\begin{tabular}{rll}
$I$  & $2$ & $I$ \\
$-I$ & $0$ & $I$%
\end{tabular}
\right) \langle I||Q||I\rangle \\
&=& 2 \sqrt{\frac{I(2I-1)}{(2I+3)(2I+1)(I+1)}} \, \langle I||Q||I \rangle .
\end{eqnarray*}
Defining $\mathbf{N}^{(1)}$ and $N_q^{(2)}$ in dimensionless form as
\begin{eqnarray*}
\mathbf{N}^{(1)} &=& {\boldsymbol \mu}/\mu_N, \\
N_q^{(2)} &=& Q_q^{(2)}/[1\,\,\mathrm{barn}],
\end{eqnarray*}
where $\mu_N$ is the nuclear magneton ($\mu_N = \frac{|e| \hbar}{2m_p c}$,
with $m_p$ being the proton mass), and the reduced matrix elements are given by
\begin{eqnarray*}
\langle I ||N^{(1)}|| I\rangle &=&\sqrt{\frac{(2I+1)(I+1)}{I}}%
\frac{\mu }{\mu_{N}}, \\
\langle I ||N^{(2)}|| I\rangle &=&\frac{1}{2}\sqrt{\frac{%
(2I+3)(2I+1)(I+1)}{I(2I-1)}}\left[ \frac{Q}{1\text{\textrm{barn}}}\right] .
\end{eqnarray*}

We define one-particle electronic tensors (in a.u.) as
\begin{eqnarray*}
T_q^{(1)} &=& - \frac{i \alpha \sqrt{2} \left( {\gamma_0 \boldsymbol \gamma} \cdot \mathbf{C}_{1q}^{(0)} (\bf \hat{r}) \right)}
{r^2} \,\, \mu_N, \\
T_q^{(2)} &=& -\frac{C_q^{(2)} (\mathbf{\hat{r}})}{r^3} \times \left[ 1\,\text{\textrm{barn}}\right] .
\end{eqnarray*}
Here $\alpha$ is the fine-structure constant, $\mathbf{C}_{1q}^{(0)}$ is a normalized
spherical harmonic, $\gamma_0$ and $\boldsymbol \gamma$ are the Dirac matrices,
and $C_q^{(2)}$ is a normalized spherical function.

\begin{thebibliography}{29}%
\makeatletter
\providecommand \@ifxundefined [1]{%
 \@ifx{#1\undefined}
}%
\providecommand \@ifnum [1]{%
 \ifnum #1\expandafter \@firstoftwo
 \else \expandafter \@secondoftwo
 \fi
}%
\providecommand \@ifx [1]{%
 \ifx #1\expandafter \@firstoftwo
 \else \expandafter \@secondoftwo
 \fi
}%
\providecommand \natexlab [1]{#1}%
\providecommand \enquote  [1]{``#1''}%
\providecommand \bibnamefont  [1]{#1}%
\providecommand \bibfnamefont [1]{#1}%
\providecommand \citenamefont [1]{#1}%
\providecommand \href@noop [0]{\@secondoftwo}%
\providecommand \href [0]{\begingroup \@sanitize@url \@href}%
\providecommand \@href[1]{\@@startlink{#1}\@@href}%
\providecommand \@@href[1]{\endgroup#1\@@endlink}%
\providecommand \@sanitize@url [0]{\catcode `\\12\catcode `\$12\catcode
  `\&12\catcode `\#12\catcode `\^12\catcode `\_12\catcode `\%12\relax}%
\providecommand \@@startlink[1]{}%
\providecommand \@@endlink[0]{}%
\providecommand \url  [0]{\begingroup\@sanitize@url \@url }%
\providecommand \@url [1]{\endgroup\@href {#1}{\urlprefix }}%
\providecommand \urlprefix  [0]{URL }%
\providecommand \Eprint [0]{\href }%
\providecommand \doibase [0]{http://dx.doi.org/}%
\providecommand \selectlanguage [0]{\@gobble}%
\providecommand \bibinfo  [0]{\@secondoftwo}%
\providecommand \bibfield  [0]{\@secondoftwo}%
\providecommand \translation [1]{[#1]}%
\providecommand \BibitemOpen [0]{}%
\providecommand \bibitemStop [0]{}%
\providecommand \bibitemNoStop [0]{.\EOS\space}%
\providecommand \EOS [0]{\spacefactor3000\relax}%
\providecommand \BibitemShut  [1]{\csname bibitem#1\endcsname}%
\let\auto@bib@innerbib\@empty
\bibitem [{\citenamefont {Rosenband}\ \emph {et~al.}(2008)\citenamefont
  {Rosenband}, \citenamefont {Hume}, \citenamefont {Schmidt}, \citenamefont
  {Chou}, \citenamefont {Brusch}, \citenamefont {Lorini}, \citenamefont
  {Oskay}, \citenamefont {Drullinger}, \citenamefont {Fortier}, \citenamefont
  {Stalnaker}, \citenamefont {Diddams}, \citenamefont {Swann}, \citenamefont
  {Newbury}, \citenamefont {Itano}, \citenamefont {Wineland},\ and\
  \citenamefont {Bergquist}}]{alpha}%
  \BibitemOpen
  \bibfield  {author} {\bibinfo {author} {\bibfnamefont {T.}~\bibnamefont
  {Rosenband}}, \bibinfo {author} {\bibfnamefont {D.~B.}\ \bibnamefont {Hume}},
  \bibinfo {author} {\bibfnamefont {P.~O.}\ \bibnamefont {Schmidt}}, \bibinfo
  {author} {\bibfnamefont {C.~W.}\ \bibnamefont {Chou}}, \bibinfo {author}
  {\bibfnamefont {A.}~\bibnamefont {Brusch}}, \bibinfo {author} {\bibfnamefont
  {L.}~\bibnamefont {Lorini}}, \bibinfo {author} {\bibfnamefont {W.~H.}\
  \bibnamefont {Oskay}}, \bibinfo {author} {\bibfnamefont {R.~E.}\ \bibnamefont
  {Drullinger}}, \bibinfo {author} {\bibfnamefont {T.~M.}\ \bibnamefont
  {Fortier}}, \bibinfo {author} {\bibfnamefont {J.~E.}\ \bibnamefont
  {Stalnaker}}, \bibinfo {author} {\bibfnamefont {S.~A.}\ \bibnamefont
  {Diddams}}, \bibinfo {author} {\bibfnamefont {W.~C.}\ \bibnamefont {Swann}},
  \bibinfo {author} {\bibfnamefont {N.~R.}\ \bibnamefont {Newbury}}, \bibinfo
  {author} {\bibfnamefont {W.~M.}\ \bibnamefont {Itano}}, \bibinfo {author}
  {\bibfnamefont {D.~J.}\ \bibnamefont {Wineland}}, \ and\ \bibinfo {author}
  {\bibfnamefont {J.~C.}\ \bibnamefont {Bergquist}},\ }\href@noop {} {\bibfield
   {journal} {\bibinfo  {journal} {Science}\ }\textbf {\bibinfo {volume}
  {319}},\ \bibinfo {pages} {1808} (\bibinfo {year} {2008})}\BibitemShut
  {NoStop}%
\bibitem [{\citenamefont {Zhang}\ \emph {et~al.}(2014)\citenamefont {Zhang},
  \citenamefont {Bishof}, \citenamefont {Bromley}, \citenamefont {Kraus},
  \citenamefont {Safronova}, \citenamefont {Zoller}, \citenamefont {Rey},\ and\
  \citenamefont {Ye}}]{SUN}%
  \BibitemOpen
  \bibfield  {author} {\bibinfo {author} {\bibfnamefont {X.}~\bibnamefont
  {Zhang}}, \bibinfo {author} {\bibfnamefont {M.}~\bibnamefont {Bishof}},
  \bibinfo {author} {\bibfnamefont {S.~L.}\ \bibnamefont {Bromley}}, \bibinfo
  {author} {\bibfnamefont {C.~V.}\ \bibnamefont {Kraus}}, \bibinfo {author}
  {\bibfnamefont {M.~S.}\ \bibnamefont {Safronova}}, \bibinfo {author}
  {\bibfnamefont {P.}~\bibnamefont {Zoller}}, \bibinfo {author} {\bibfnamefont
  {A.~M.}\ \bibnamefont {Rey}}, \ and\ \bibinfo {author} {\bibfnamefont
  {J.}~\bibnamefont {Ye}},\ }\href@noop {} {\bibfield  {journal} {\bibinfo
  {journal} {Science}\ }\textbf {\bibinfo {volume} {345}},\ \bibinfo {pages}
  {1467} (\bibinfo {year} {2014})}\BibitemShut {NoStop}%
\bibitem [{\citenamefont {Martin}\ \emph {et~al.}(2013)\citenamefont {Martin},
  \citenamefont {Bishof}, \citenamefont {Swallows}, \citenamefont {Zhang},
  \citenamefont {Benko}, \citenamefont {von Stecher}, \citenamefont {Gorshkov},
  \citenamefont {Rey},\ and\ \citenamefont {Ye}}]{QMBYe}%
  \BibitemOpen
  \bibfield  {author} {\bibinfo {author} {\bibfnamefont {M.~J.}\ \bibnamefont
  {Martin}}, \bibinfo {author} {\bibfnamefont {M.}~\bibnamefont {Bishof}},
  \bibinfo {author} {\bibfnamefont {M.~D.}\ \bibnamefont {Swallows}}, \bibinfo
  {author} {\bibfnamefont {X.}~\bibnamefont {Zhang}}, \bibinfo {author}
  {\bibfnamefont {C.}~\bibnamefont {Benko}}, \bibinfo {author} {\bibfnamefont
  {J.}~\bibnamefont {von Stecher}}, \bibinfo {author} {\bibfnamefont {A.~V.}\
  \bibnamefont {Gorshkov}}, \bibinfo {author} {\bibfnamefont {A.~M.}\
  \bibnamefont {Rey}}, \ and\ \bibinfo {author} {\bibfnamefont
  {J.}~\bibnamefont {Ye}},\ }\href@noop {} {\bibfield  {journal} {\bibinfo
  {journal} {Science}\ }\textbf {\bibinfo {volume} {314}},\ \bibinfo {pages}
  {632} (\bibinfo {year} {2013})}\BibitemShut {NoStop}%
\bibitem [{\citenamefont {Chou}\ \emph {et~al.}(2010)\citenamefont {Chou},
  \citenamefont {Hume}, \citenamefont {Koelemeij}, \citenamefont {Wineland},\
  and\ \citenamefont {Rosenband}}]{AlIon}%
  \BibitemOpen
  \bibfield  {author} {\bibinfo {author} {\bibfnamefont {C.~W.}\ \bibnamefont
  {Chou}}, \bibinfo {author} {\bibfnamefont {D.~B.}\ \bibnamefont {Hume}},
  \bibinfo {author} {\bibfnamefont {J.~C.~J.}\ \bibnamefont {Koelemeij}},
  \bibinfo {author} {\bibfnamefont {D.~J.}\ \bibnamefont {Wineland}}, \ and\
  \bibinfo {author} {\bibfnamefont {T.}~\bibnamefont {Rosenband}},\ }\href@noop
  {} {\bibfield  {journal} {\bibinfo  {journal} {Phys. Rev. Lett.}\ }\textbf
  {\bibinfo {volume} {104}},\ \bibinfo {pages} {070802} (\bibinfo {year}
  {2010})}\BibitemShut {NoStop}%
\bibitem [{\citenamefont {Bloom}\ \emph {et~al.}(2014)\citenamefont {Bloom},
  \citenamefont {Nicholson}, \citenamefont {Williams}, \citenamefont
  {Campbell}, \citenamefont {Bishof}, \citenamefont {Zhang}, \citenamefont
  {Zhang}, \citenamefont {Bromley},\ and\ \citenamefont {Ye}}]{SrYe}%
  \BibitemOpen
  \bibfield  {author} {\bibinfo {author} {\bibfnamefont {B.~J.}\ \bibnamefont
  {Bloom}}, \bibinfo {author} {\bibfnamefont {T.~L.}\ \bibnamefont
  {Nicholson}}, \bibinfo {author} {\bibfnamefont {J.~R.}\ \bibnamefont
  {Williams}}, \bibinfo {author} {\bibfnamefont {S.~L.}\ \bibnamefont
  {Campbell}}, \bibinfo {author} {\bibfnamefont {M.}~\bibnamefont {Bishof}},
  \bibinfo {author} {\bibfnamefont {X.}~\bibnamefont {Zhang}}, \bibinfo
  {author} {\bibfnamefont {W.}~\bibnamefont {Zhang}}, \bibinfo {author}
  {\bibfnamefont {S.~L.}\ \bibnamefont {Bromley}}, \ and\ \bibinfo {author}
  {\bibfnamefont {J.}~\bibnamefont {Ye}},\ }\href@noop {} {\bibfield  {journal}
  {\bibinfo  {journal} {Nature.}\ }\textbf {\bibinfo {volume} {506}},\ \bibinfo
  {pages} {71} (\bibinfo {year} {2014})}\BibitemShut {NoStop}%
\bibitem [{\citenamefont {Stalnaker}\ \emph {et~al.}(2007)\citenamefont
  {Stalnaker}, \citenamefont {Diddams}, \citenamefont {Fortier}, \citenamefont
  {Kim}, \citenamefont {Hollberg}, \citenamefont {Bergquist}, \citenamefont
  {Itano}, \citenamefont {Delany}, \citenamefont {Lorini}, \citenamefont
  {Oskay}, \citenamefont {Heavner}, \citenamefont {Jefferts}, \citenamefont
  {Levi}, \citenamefont {Parker},\ and\ \citenamefont {Shirley}}]{HgIon}%
  \BibitemOpen
  \bibfield  {author} {\bibinfo {author} {\bibfnamefont {J.~E.}\ \bibnamefont
  {Stalnaker}}, \bibinfo {author} {\bibfnamefont {S.}~\bibnamefont {Diddams}},
  \bibinfo {author} {\bibfnamefont {T.}~\bibnamefont {Fortier}}, \bibinfo
  {author} {\bibfnamefont {K.}~\bibnamefont {Kim}}, \bibinfo {author}
  {\bibfnamefont {L.}~\bibnamefont {Hollberg}}, \bibinfo {author}
  {\bibfnamefont {J.}~\bibnamefont {Bergquist}}, \bibinfo {author}
  {\bibfnamefont {W.}~\bibnamefont {Itano}}, \bibinfo {author} {\bibfnamefont
  {M.}~\bibnamefont {Delany}}, \bibinfo {author} {\bibfnamefont
  {L.}~\bibnamefont {Lorini}}, \bibinfo {author} {\bibfnamefont
  {W.}~\bibnamefont {Oskay}}, \bibinfo {author} {\bibfnamefont
  {T.}~\bibnamefont {Heavner}}, \bibinfo {author} {\bibfnamefont
  {S.}~\bibnamefont {Jefferts}}, \bibinfo {author} {\bibfnamefont
  {F.}~\bibnamefont {Levi}}, \bibinfo {author} {\bibfnamefont {T.}~\bibnamefont
  {Parker}}, \ and\ \bibinfo {author} {\bibfnamefont {J.}~\bibnamefont
  {Shirley}},\ }\href@noop {} {\bibfield  {journal} {\bibinfo  {journal} {Appl.
  Phys. B}\ }\textbf {\bibinfo {volume} {89}},\ \bibinfo {pages} {167}
  (\bibinfo {year} {2007})}\BibitemShut {NoStop}%
\bibitem [{\citenamefont {Dube}\ \emph {et~al.}(2013)\citenamefont {Dube},
  \citenamefont {Madej}, \citenamefont {Zhou},\ and\ \citenamefont
  {Bernard}}]{SrIon}%
  \BibitemOpen
  \bibfield  {author} {\bibinfo {author} {\bibfnamefont {P.}~\bibnamefont
  {Dube}}, \bibinfo {author} {\bibfnamefont {A.~A.}\ \bibnamefont {Madej}},
  \bibinfo {author} {\bibfnamefont {Z.}~\bibnamefont {Zhou}}, \ and\ \bibinfo
  {author} {\bibfnamefont {J.~E.}\ \bibnamefont {Bernard}},\ }\href@noop {}
  {\bibfield  {journal} {\bibinfo  {journal} {Phys. Rev. A.}\ }\textbf
  {\bibinfo {volume} {87}},\ \bibinfo {pages} {023806} (\bibinfo {year}
  {2013})}\BibitemShut {NoStop}%
\bibitem [{\citenamefont {Huntemann}\ \emph {et~al.}(2012)\citenamefont
  {Huntemann}, \citenamefont {Okhapkin}, \citenamefont {Lipphardt},
  \citenamefont {Weyers}, \citenamefont {Tamm},\ and\ \citenamefont
  {Peik}}]{YbIon}%
  \BibitemOpen
  \bibfield  {author} {\bibinfo {author} {\bibfnamefont {N.}~\bibnamefont
  {Huntemann}}, \bibinfo {author} {\bibfnamefont {M.}~\bibnamefont {Okhapkin}},
  \bibinfo {author} {\bibfnamefont {B.}~\bibnamefont {Lipphardt}}, \bibinfo
  {author} {\bibfnamefont {S.}~\bibnamefont {Weyers}}, \bibinfo {author}
  {\bibfnamefont {C.}~\bibnamefont {Tamm}}, \ and\ \bibinfo {author}
  {\bibfnamefont {E.}~\bibnamefont {Peik}},\ }\href@noop {} {\bibfield
  {journal} {\bibinfo  {journal} {Phys. Rev. Lett.}\ }\textbf {\bibinfo
  {volume} {108}},\ \bibinfo {pages} {090801} (\bibinfo {year}
  {2012})}\BibitemShut {NoStop}%
\bibitem [{\citenamefont {Wang}\ \emph {et~al.}(2007)\citenamefont {Wang},
  \citenamefont {Dumke}, \citenamefont {Liu}, \citenamefont {Stejskal},
  \citenamefont {Zhao}, \citenamefont {Zhang}, \citenamefont {Lu},
  \citenamefont {Wang}, \citenamefont {Becker},\ and\ \citenamefont
  {Walther}}]{InIon}%
  \BibitemOpen
  \bibfield  {author} {\bibinfo {author} {\bibfnamefont {Y.~H.}\ \bibnamefont
  {Wang}}, \bibinfo {author} {\bibfnamefont {R.}~\bibnamefont {Dumke}},
  \bibinfo {author} {\bibfnamefont {T.}~\bibnamefont {Liu}}, \bibinfo {author}
  {\bibfnamefont {A.}~\bibnamefont {Stejskal}}, \bibinfo {author}
  {\bibfnamefont {Y.}~\bibnamefont {Zhao}}, \bibinfo {author} {\bibfnamefont
  {J.}~\bibnamefont {Zhang}}, \bibinfo {author} {\bibfnamefont
  {Z.}~\bibnamefont {Lu}}, \bibinfo {author} {\bibfnamefont {L.~J.}\
  \bibnamefont {Wang}}, \bibinfo {author} {\bibfnamefont {T.}~\bibnamefont
  {Becker}}, \ and\ \bibinfo {author} {\bibfnamefont {H.}~\bibnamefont
  {Walther}},\ }\href@noop {} {\bibfield  {journal} {\bibinfo  {journal} {Opt.
  Commun.}\ }\textbf {\bibinfo {volume} {273}},\ \bibinfo {pages} {526}
  (\bibinfo {year} {2007})}\BibitemShut {NoStop}%
\bibitem [{\citenamefont {Barber}\ \emph {et~al.}(2006)\citenamefont {Barber},
  \citenamefont {Hoyt}, \citenamefont {Oates}, \citenamefont {Hollberg},
  \citenamefont {Taichenachev},\ and\ \citenamefont {Yudin}}]{YbForbidden}%
  \BibitemOpen
  \bibfield  {author} {\bibinfo {author} {\bibfnamefont {Z.~W.}\ \bibnamefont
  {Barber}}, \bibinfo {author} {\bibfnamefont {C.~W.}\ \bibnamefont {Hoyt}},
  \bibinfo {author} {\bibfnamefont {C.~W.}\ \bibnamefont {Oates}}, \bibinfo
  {author} {\bibfnamefont {L.}~\bibnamefont {Hollberg}}, \bibinfo {author}
  {\bibfnamefont {A.~V.}\ \bibnamefont {Taichenachev}}, \ and\ \bibinfo
  {author} {\bibfnamefont {V.~I.}\ \bibnamefont {Yudin}},\ }\href@noop {}
  {\bibfield  {journal} {\bibinfo  {journal} {Phys. Rev. Lett.}\ }\textbf
  {\bibinfo {volume} {96}},\ \bibinfo {pages} {083002} (\bibinfo {year}
  {2006})}\BibitemShut {NoStop}%
\bibitem [{\citenamefont {Ludlow}\ \emph {et~al.}(2015)\citenamefont {Ludlow},
  \citenamefont {Boyd}, \citenamefont {Ye}, \citenamefont {Peik},\ and\
  \citenamefont {Schmidt}}]{RMP}%
  \BibitemOpen
  \bibfield  {author} {\bibinfo {author} {\bibfnamefont {A.~D.}\ \bibnamefont
  {Ludlow}}, \bibinfo {author} {\bibfnamefont {M.~M.}\ \bibnamefont {Boyd}},
  \bibinfo {author} {\bibfnamefont {J.}~\bibnamefont {Ye}}, \bibinfo {author}
  {\bibfnamefont {E.}~\bibnamefont {Peik}}, \ and\ \bibinfo {author}
  {\bibfnamefont {P.}~\bibnamefont {Schmidt}},\ }\href@noop {} {\bibfield
  {journal} {\bibinfo  {journal} {Rev. Mod. Phys.}\ }\textbf {\bibinfo {volume}
  {87}},\ \bibinfo {pages} {637} (\bibinfo {year} {2015})}\BibitemShut
  {NoStop}%
\bibitem [{\citenamefont {Hinkley}\ \emph {et~al.}(2013)\citenamefont
  {Hinkley}, \citenamefont {Sherman}, \citenamefont {Phillips}, \citenamefont
  {Schioppo}, \citenamefont {Lemke}, \citenamefont {Beloy}, \citenamefont
  {Pizzocaro}, \citenamefont {Oates},\ and\ \citenamefont {Ludlow}}]{Ludlow}%
  \BibitemOpen
  \bibfield  {author} {\bibinfo {author} {\bibfnamefont {N.}~\bibnamefont
  {Hinkley}}, \bibinfo {author} {\bibfnamefont {J.~A.}\ \bibnamefont
  {Sherman}}, \bibinfo {author} {\bibfnamefont {N.~B.}\ \bibnamefont
  {Phillips}}, \bibinfo {author} {\bibfnamefont {M.}~\bibnamefont {Schioppo}},
  \bibinfo {author} {\bibfnamefont {N.~D.}\ \bibnamefont {Lemke}}, \bibinfo
  {author} {\bibfnamefont {K.}~\bibnamefont {Beloy}}, \bibinfo {author}
  {\bibfnamefont {M.}~\bibnamefont {Pizzocaro}}, \bibinfo {author}
  {\bibfnamefont {C.~W.}\ \bibnamefont {Oates}}, \ and\ \bibinfo {author}
  {\bibfnamefont {A.~D.}\ \bibnamefont {Ludlow}},\ }\href@noop {} {\bibfield
  {journal} {\bibinfo  {journal} {Science}\ }\textbf {\bibinfo {volume}
  {341}},\ \bibinfo {pages} {1215} (\bibinfo {year} {2013})}\BibitemShut
  {NoStop}%
\bibitem [{\citenamefont {Nicholson}\ \emph {et~al.}(2015)\citenamefont
  {Nicholson}, \citenamefont {Campbell}, \citenamefont {Hutson}, \citenamefont
  {Marti}, \citenamefont {Bloom}, \citenamefont {McNally}, \citenamefont
  {Zhang}, \citenamefont {Barrett}, \citenamefont {Safronova}, \citenamefont
  {Strouse}, \citenamefont {Tew},\ and\ \citenamefont {Ye}}]{SrYe2}%
  \BibitemOpen
  \bibfield  {author} {\bibinfo {author} {\bibfnamefont {T.~L.}\ \bibnamefont
  {Nicholson}}, \bibinfo {author} {\bibfnamefont {S.~L.}\ \bibnamefont
  {Campbell}}, \bibinfo {author} {\bibfnamefont {R.~B.}\ \bibnamefont
  {Hutson}}, \bibinfo {author} {\bibfnamefont {G.~E.}\ \bibnamefont {Marti}},
  \bibinfo {author} {\bibfnamefont {B.~J.}\ \bibnamefont {Bloom}}, \bibinfo
  {author} {\bibfnamefont {R.~L.}\ \bibnamefont {McNally}}, \bibinfo {author}
  {\bibfnamefont {W.}~\bibnamefont {Zhang}}, \bibinfo {author} {\bibfnamefont
  {M.~D.}\ \bibnamefont {Barrett}}, \bibinfo {author} {\bibfnamefont {M.~S.}\
  \bibnamefont {Safronova}}, \bibinfo {author} {\bibfnamefont {G.}~\bibnamefont
  {Strouse}}, \bibinfo {author} {\bibfnamefont {W.~L.}\ \bibnamefont {Tew}}, \
  and\ \bibinfo {author} {\bibfnamefont {J.}~\bibnamefont {Ye}},\ }\href@noop
  {} {\bibfield  {journal} {\bibinfo  {journal} {Nature. Comm.}\ }\textbf
  {\bibinfo {volume} {6}},\ \bibinfo {pages} {6896} (\bibinfo {year}
  {2015})}\BibitemShut {NoStop}%
\bibitem [{\citenamefont {Barrett}(2015)}]{MDB1}%
  \BibitemOpen
  \bibfield  {author} {\bibinfo {author} {\bibfnamefont {M.~D.}\ \bibnamefont
  {Barrett}},\ }\href@noop {} {\bibfield  {journal} {\bibinfo  {journal} {New
  Jour. Phys.}\ }\textbf {\bibinfo {volume} {17}},\ \bibinfo {pages} {053024}
  (\bibinfo {year} {2015})}\BibitemShut {NoStop}%
\bibitem [{\citenamefont {Arnold}\ \emph {et~al.}(2015)\citenamefont {Arnold},
  \citenamefont {Hajiyev}, \citenamefont {Paez}, \citenamefont {Lee},
  \citenamefont {Barrett},\ and\ \citenamefont {Bollinger}}]{MDB2}%
  \BibitemOpen
  \bibfield  {author} {\bibinfo {author} {\bibfnamefont {K.}~\bibnamefont
  {Arnold}}, \bibinfo {author} {\bibfnamefont {E.}~\bibnamefont {Hajiyev}},
  \bibinfo {author} {\bibfnamefont {E.}~\bibnamefont {Paez}}, \bibinfo {author}
  {\bibfnamefont {C.~H.}\ \bibnamefont {Lee}}, \bibinfo {author} {\bibfnamefont
  {M.~D.}\ \bibnamefont {Barrett}}, \ and\ \bibinfo {author} {\bibfnamefont
  {J.}~\bibnamefont {Bollinger}},\ }\href@noop {} {\bibfield  {journal}
  {\bibinfo  {journal} {Phys. Rev. A.}\ }\textbf {\bibinfo {volume} {92}},\
  \bibinfo {pages} {032108} (\bibinfo {year} {2015})}\BibitemShut {NoStop}%
\bibitem [{\citenamefont {Kozlov}\ \emph {et~al.}(2014)\citenamefont {Kozlov},
  \citenamefont {Dzuba},\ and\ \citenamefont {Flambaum}}]{Dzuba}%
  \BibitemOpen
  \bibfield  {author} {\bibinfo {author} {\bibfnamefont {A.}~\bibnamefont
  {Kozlov}}, \bibinfo {author} {\bibfnamefont {V.~A.}\ \bibnamefont {Dzuba}}, \
  and\ \bibinfo {author} {\bibfnamefont {V.~V.}\ \bibnamefont {Flambaum}},\
  }\href@noop {} {\bibfield  {journal} {\bibinfo  {journal} {Phys. Rev. A.}\
  }\textbf {\bibinfo {volume} {90}},\ \bibinfo {pages} {042505} (\bibinfo
  {year} {2014})}\BibitemShut {NoStop}%
\bibitem [{\citenamefont {Chuah}\ \emph {et~al.}(2011)\citenamefont {Chuah},
  \citenamefont {Lewty},\ and\ \citenamefont {Barrett}}]{BoonLeng2}%
  \BibitemOpen
  \bibfield  {author} {\bibinfo {author} {\bibfnamefont {B.~L.}\ \bibnamefont
  {Chuah}}, \bibinfo {author} {\bibfnamefont {N.~C.}\ \bibnamefont {Lewty}}, \
  and\ \bibinfo {author} {\bibfnamefont {M.~D.}\ \bibnamefont {Barrett}},\
  }\href@noop {} {\bibfield  {journal} {\bibinfo  {journal} {Phys. Rev. A}\
  }\textbf {\bibinfo {volume} {84}},\ \bibinfo {pages} {013411} (\bibinfo
  {year} {2011})}\BibitemShut {NoStop}%
\bibitem [{\citenamefont {Lewty}\ \emph {et~al.}(2012)\citenamefont {Lewty},
  \citenamefont {Chuah}, \citenamefont {Cazan}, \citenamefont {Sahoo},\ and\
  \citenamefont {Barrett}}]{Nick}%
  \BibitemOpen
  \bibfield  {author} {\bibinfo {author} {\bibfnamefont {N.~C.}\ \bibnamefont
  {Lewty}}, \bibinfo {author} {\bibfnamefont {B.~L.}\ \bibnamefont {Chuah}},
  \bibinfo {author} {\bibfnamefont {R.}~\bibnamefont {Cazan}}, \bibinfo
  {author} {\bibfnamefont {B.~K.}\ \bibnamefont {Sahoo}}, \ and\ \bibinfo
  {author} {\bibfnamefont {M.~D.}\ \bibnamefont {Barrett}},\ }\href@noop {}
  {\bibfield  {journal} {\bibinfo  {journal} {Optics Express}\ }\textbf
  {\bibinfo {volume} {20}},\ \bibinfo {pages} {21379} (\bibinfo {year}
  {2012})}\BibitemShut {NoStop}%
\bibitem [{\citenamefont {Myerson}\ \emph {et~al.}(2008)\citenamefont
  {Myerson}, \citenamefont {Szwer}, \citenamefont {Webster}, \citenamefont
  {Allcock}, \citenamefont {Curtis}, \citenamefont {Imreh}, \citenamefont
  {Sherman}, \citenamefont {Stacey}, \citenamefont {Steane},\ and\
  \citenamefont {Lucas}}]{Lucas}%
  \BibitemOpen
  \bibfield  {author} {\bibinfo {author} {\bibfnamefont {A.~H.}\ \bibnamefont
  {Myerson}}, \bibinfo {author} {\bibfnamefont {D.~J.}\ \bibnamefont {Szwer}},
  \bibinfo {author} {\bibfnamefont {S.~C.}\ \bibnamefont {Webster}}, \bibinfo
  {author} {\bibfnamefont {D.~T.~C.}\ \bibnamefont {Allcock}}, \bibinfo
  {author} {\bibfnamefont {M.~J.}\ \bibnamefont {Curtis}}, \bibinfo {author}
  {\bibfnamefont {G.}~\bibnamefont {Imreh}}, \bibinfo {author} {\bibfnamefont
  {J.~A.}\ \bibnamefont {Sherman}}, \bibinfo {author} {\bibfnamefont {D.~N.}\
  \bibnamefont {Stacey}}, \bibinfo {author} {\bibfnamefont {A.~M.}\
  \bibnamefont {Steane}}, \ and\ \bibinfo {author} {\bibfnamefont {D.~M.}\
  \bibnamefont {Lucas}},\ }\href@noop {} {\bibfield  {journal} {\bibinfo
  {journal} {Phys Rev Lett}\ }\textbf {\bibinfo {volume} {100}},\ \bibinfo
  {pages} {200502} (\bibinfo {year} {2008})}\BibitemShut {NoStop}%
\bibitem [{\citenamefont {{Safronova}}\ \emph {et~al.}(2009)\citenamefont
  {{Safronova}}, \citenamefont {{Kozlov}}, \citenamefont {{Johnson}},\ and\
  \citenamefont {{Jiang}}}]{SafKozJoh09}%
  \BibitemOpen
  \bibfield  {author} {\bibinfo {author} {\bibfnamefont {M.~S.}\ \bibnamefont
  {{Safronova}}}, \bibinfo {author} {\bibfnamefont {M.~G.}\ \bibnamefont
  {{Kozlov}}}, \bibinfo {author} {\bibfnamefont {W.~R.}\ \bibnamefont
  {{Johnson}}}, \ and\ \bibinfo {author} {\bibfnamefont {D.}~\bibnamefont
  {{Jiang}}},\ }\href@noop {} {\bibfield  {journal} {\bibinfo  {journal} {Phys.
  Rev. A}\ }\textbf {\bibinfo {volume} {80}},\ \bibinfo {pages} {012516}
  (\bibinfo {year} {2009})}\BibitemShut {NoStop}%
\bibitem [{\citenamefont {Dzuba}\ \emph {et~al.}(1996)\citenamefont {Dzuba},
  \citenamefont {Flambaum},\ and\ \citenamefont {Kozlov}}]{DzuFlaKoz96}%
  \BibitemOpen
  \bibfield  {author} {\bibinfo {author} {\bibfnamefont {V.~A.}\ \bibnamefont
  {Dzuba}}, \bibinfo {author} {\bibfnamefont {V.~V.}\ \bibnamefont {Flambaum}},
  \ and\ \bibinfo {author} {\bibfnamefont {M.~G.}\ \bibnamefont {Kozlov}},\
  }\href@noop {} {\bibfield  {journal} {\bibinfo  {journal} {Phys.\ Rev.\ A}\
  }\textbf {\bibinfo {volume} {54}},\ \bibinfo {pages} {3948} (\bibinfo {year}
  {1996})}\BibitemShut {NoStop}%
\bibitem [{\citenamefont {Dzuba}\ \emph {et~al.}(1998)\citenamefont {Dzuba},
  \citenamefont {Kozlov}, \citenamefont {Porsev},\ and\ \citenamefont
  {Flambaum}}]{DzuKozPor98}%
  \BibitemOpen
  \bibfield  {author} {\bibinfo {author} {\bibfnamefont {V.~A.}\ \bibnamefont
  {Dzuba}}, \bibinfo {author} {\bibfnamefont {M.~G.}\ \bibnamefont {Kozlov}},
  \bibinfo {author} {\bibfnamefont {S.~G.}\ \bibnamefont {Porsev}}, \ and\
  \bibinfo {author} {\bibfnamefont {V.~V.}\ \bibnamefont {Flambaum}},\
  }\href@noop {} {\bibfield  {journal} {\bibinfo  {journal} {Zh. \ Eksp. \
  Teor. \ Fiz.}\ }\textbf {\bibinfo {volume} {114}},\ \bibinfo {pages} {1636}
  (\bibinfo {year} {1998})},\ \bibinfo {note} {[Sov. \ Phys.--JETP {\bf 87}
  885, (1998)]}\BibitemShut {NoStop}%
\bibitem [{Ral()}]{RalKraRea11}%
  \BibitemOpen
  \href@noop {} {}\bibinfo {note} {Yu.~Ralchenko, A.~Kramida, J.~Reader, and
  the NIST ASD Team (2011). NIST Atomic Spectra Database (version 4.1).
  Available at http://physics.nist.gov/asd. National Institute of Standards and
  Technology, Gaithersburg, MD.}\BibitemShut {Stop}%
\bibitem [{\citenamefont {Sternheimer}(1950)}]{Ste50}%
  \BibitemOpen
  \bibfield  {author} {\bibinfo {author} {\bibfnamefont {R.~M.}\ \bibnamefont
  {Sternheimer}},\ }\href@noop {} {\bibfield  {journal} {\bibinfo  {journal}
  {Phys. Rev.}\ }\textbf {\bibinfo {volume} {80}},\ \bibinfo {pages} {102}
  (\bibinfo {year} {1950})}\BibitemShut {NoStop}%
\bibitem [{\citenamefont {Dalgarno}\ and\ \citenamefont
  {Lewis}(1955)}]{DalLew55}%
  \BibitemOpen
  \bibfield  {author} {\bibinfo {author} {\bibfnamefont {A.}~\bibnamefont
  {Dalgarno}}\ and\ \bibinfo {author} {\bibfnamefont {J.~T.}\ \bibnamefont
  {Lewis}},\ }\href@noop {} {\bibfield  {journal} {\bibinfo  {journal} {Proc.
  R. Soc. London, Ser. A}\ }\textbf {\bibinfo {volume} {223}},\ \bibinfo
  {pages} {70} (\bibinfo {year} {1955})}\BibitemShut {NoStop}%
\bibitem [{\citenamefont {{Kozlov}}\ and\ \citenamefont
  {{Porsev}}(1999)}]{KozPor99a}%
  \BibitemOpen
  \bibfield  {author} {\bibinfo {author} {\bibfnamefont {M.~G.}\ \bibnamefont
  {{Kozlov}}}\ and\ \bibinfo {author} {\bibfnamefont {S.~G.}\ \bibnamefont
  {{Porsev}}},\ }\href {\doibase 10.1007/s100530050229} {\bibfield  {journal}
  {\bibinfo  {journal} {Eur.~Phys.~J.~D}\ }\textbf {\bibinfo {volume} {5}},\
  \bibinfo {pages} {59} (\bibinfo {year} {1999})}\BibitemShut {NoStop}%
\bibitem [{\citenamefont {{Safronova}}\ \emph {et~al.}(2013)\citenamefont
  {{Safronova}}, \citenamefont {{Porsev}}, \citenamefont {{Safronova}},
  \citenamefont {{Kozlov}},\ and\ \citenamefont {{Clark}}}]{SafPorSaf13}%
  \BibitemOpen
  \bibfield  {author} {\bibinfo {author} {\bibfnamefont {M.~S.}\ \bibnamefont
  {{Safronova}}}, \bibinfo {author} {\bibfnamefont {S.~G.}\ \bibnamefont
  {{Porsev}}}, \bibinfo {author} {\bibfnamefont {U.~I.}\ \bibnamefont
  {{Safronova}}}, \bibinfo {author} {\bibfnamefont {M.~G.}\ \bibnamefont
  {{Kozlov}}}, \ and\ \bibinfo {author} {\bibfnamefont {C.~W.}\ \bibnamefont
  {{Clark}}},\ }\href {\doibase 10.1103/PhysRevA.87.012509} {\bibfield
  {journal} {\bibinfo  {journal} {Phys. Rev. A}\ }\textbf {\bibinfo {volume}
  {87}},\ \bibinfo {eid} {012509} (\bibinfo {year} {2013})}\BibitemShut
  {NoStop}%
\bibitem [{\citenamefont {Brenner}\ \emph {et~al.}(1985)\citenamefont
  {Brenner}, \citenamefont {U\"{u}ttgenbach}, \citenamefont {Rupprecht},\ and\
  \citenamefont {Tra\"{a}her}}]{NuclearDipole}%
  \BibitemOpen
  \bibfield  {author} {\bibinfo {author} {\bibfnamefont {T.}~\bibnamefont
  {Brenner}}, \bibinfo {author} {\bibfnamefont {S.}~\bibnamefont
  {U\"{u}ttgenbach}}, \bibinfo {author} {\bibfnamefont {W.}~\bibnamefont
  {Rupprecht}}, \ and\ \bibinfo {author} {\bibfnamefont {F.}~\bibnamefont
  {Tra\"{a}her}},\ }\href@noop {} {\bibfield  {journal} {\bibinfo  {journal}
  {Nuclear Physics A}\ }\textbf {\bibinfo {volume} {440}},\ \bibinfo {pages}
  {407} (\bibinfo {year} {1985})}\BibitemShut {NoStop}%
\bibitem [{\citenamefont {Pyykk\"{o}}(2001)}]{NuclearQuad}%
  \BibitemOpen
  \bibfield  {author} {\bibinfo {author} {\bibfnamefont {P.}~\bibnamefont
  {Pyykk\"{o}}},\ }\href@noop {} {\bibfield  {journal} {\bibinfo  {journal}
  {Molecular Physics}\ }\textbf {\bibinfo {volume} {99}},\ \bibinfo {pages}
  {1617} (\bibinfo {year} {2001})}\BibitemShut {NoStop}%
\end{thebibliography}
%
\end{document}